\documentclass[a4paper,fleqn,usenatbib]{mnras}
\usepackage{newtxtext,newtxmath}
\usepackage[T1]{fontenc}
\usepackage{ae,aecompl}
\usepackage{graphicx}   
\usepackage{amsmath}    
\usepackage{amssymb}    

\title[Modelling the KIC8462852 light curves]
{Modelling the KIC8462852 light curves: 
compatibility of the dips and secular dimming with an exocomet interpretation}
\author[M. C. Wyatt et al.]
  {M. C. Wyatt$^1$\thanks{Email: wyatt@ast.cam.ac.uk},
  R. van Lieshout$^1$,
  G. M. Kennedy$^1$,
  T. S. Boyajian$^2$\\
  $^1$ Institute of Astronomy, University of Cambridge, Madingley Road,
  Cambridge CB3 0HA, UK \\
  $^2$ Department of Physics \& Astronomy, Louisiana State University, 202 Nicholsom Hall, Baton Rouge, LA 70803, USA
}

\pubyear{2017}

\begin{document}
\label{firstpage}
\pagerange{\pageref{firstpage}--\pageref{lastpage}}
\maketitle

\begin{abstract}
This paper shows how the dips and secular dimming in the KIC8462852 light curve can
originate in circumstellar material distributed around a single elliptical orbit (e.g., exocomets).
The expected thermal emission and wavelength dependent dimming is derived for different orbital parameters and geometries,
including dust that is optically thick to stellar radiation, and for a size distribution of dust with realistic optical properties.
We first consider dust distributed evenly around the orbit, then show how to derive its uneven distribution
from the optical light curve and to predict light curves at different wavelengths.
The fractional luminosity of an even distribution is approximately the level of dimming times
stellar radius divided by distance from the star at transit.
Non-detection of dust thermal emission for KIC8462852 thus provides a lower limit on the transit distance to
complement the 0.6\,au upper limit imposed by 0.4\,day dips.
Unless the dust distribution is optically thick, the putative 16\% century-long secular dimming must have
disappeared before the {\it WISE} 12\,$\mu$m measurement in 2010,
and subsequent 4.5\,$\mu$m observations require transits at $>0.05$\,au.
However, self-absorption of thermal emission removes these constraints for opaque dust distributions.
The passage of dust clumps through pericentre is predicted to cause infrared brightening lasting 10s of days
and dimming during transit, such that total flux received decreases at wavelengths $<5$\,$\mu$m, but increases
to potentially detectable levels at longer wavelengths.
We suggest that lower dimming levels than seen for KIC8462852 are more common in the Galactic population
and may be detected in future transit surveys.
\end{abstract}

\begin{keywords}
  comets: general --
  infrared: planetary systems --
  circumstellar matter --
  planetary systems --
  stars: variables: general.
\end{keywords}

\section{Introduction}
\label{s:intro}
One of the most intriguing discoveries made by the {\it Kepler} Mission was of the aperiodic dimming observed toward
the star KIC8462852 \citep{Boyajian2016}.
This seemingly normal main sequence F star was found to exhibit dips in the light-cuve during which the
stellar flux dropped by up to 20\% in events that lasted from a fraction of a day to several days.
Unlike the planetary transit signals discovered by {\it Kepler} \citep{Rowe2015}, during which the drop in stellar flux is
moreover much smaller,
or the transiting dusty material of WD1145+017 \citep{Vanderburg2015}, these dips did not obviously repeat during the
{\it Kepler} survey which had a duration of 1500~days.
Furthermore, unlike the dippers seen around young stars \citep{MoralesCalderon2011,Ansdell2016}, no excess infrared emission above
the stellar photosphere was detected from circumstellar debris. 

\begin{figure*}
  \begin{center}
    \vspace{-0.1in}
    \begin{tabular}{c}
      \includegraphics[width=2\columnwidth]{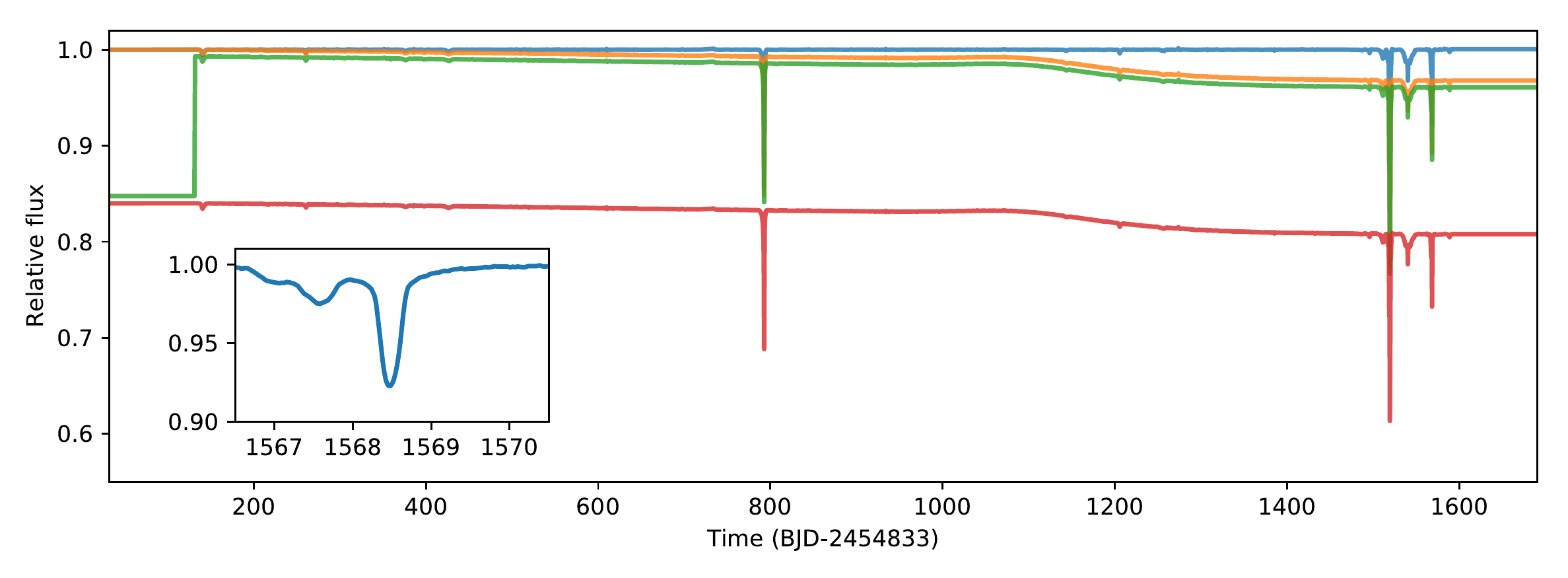}
    \end{tabular}
    \vspace{-0.15in}
    \caption{The {\it Kepler} light curve of KIC8462852 with four different interpretations of the secular
    dimming that are considered in this paper:
    the blue line has no secular dimming \citep[as presented in][]{Boyajian2016},
    the orange line includes the long-term dimming seen in the {\it Kepler} data \citep[as presented in][]{Montet2016},
    the red line assumes that the putative century long dimming \citep{Schaefer2016} is real and
    continued to provide 16\% dimming at the start of the {\it Kepler} observations
    \citep[using data from][]{Montet2016},
    the green line assumes that the putative century long dimming is real but the
    level of dimming had returned to normal before the start of the {\it Kepler} observations.
    The lines are slightly offset vertically for clarity (the light curves without this offset are shown
    in Fig.~\ref{fig:irvt}).
    The inset shows the shortest duration dip present in the light curve.
    }
   \label{fig:lczoom}
  \end{center}
\end{figure*}

\citet{Boyajian2016} showed that, if the dips come from dusty material on a circumstellar orbit, constraints
from the light curve itself pointed to an eccentric orbit that could be oriented with its pericentre direction
along our line of sight.
This implies that multiple dust clumps could be distributed along the orbit and transit in front of the star as they
pass through pericentre.
The distribution of these clumps would be reminiscent of that resulting from the break-up of comets in the Solar System
\citep{Weaver1995,Movshovitz2012}.
This analogy was reinforced by \citet{Bodman2016} who modelled the passage of comets in front of the
star, showing that the light curve could be reproduced, albeit that assuming dust production rates 
scaled from Solar System comets suggests a rather high total mass in comets on this orbit.
\citet{Neslusan2017} came to a similar conclusion, though modelling the clumps as originating in more massive
parent bodies orbiting the star.

Subsequent infrared observations further confirmed and quantified the lack of excess emission
\citep{Marengo2015,Lisse2015,Thompson2016},
while analysis of archival plates found that the light curve of this star was even more interesting
than originally thought.
Evidence was found that the light from the star has been dimming over
century timescales, with a 16\% decrease in brightness from 1890 to 1980 \citep{Schaefer2016},
the so-called {\it secular dimming}.
This secular dimming remains controversial, since analysis of the same archival data
found no evidence for the dimming \citep{Hippke2016}, and a different set of archival data also
found no evidence for secular dimming \citep{Hippke2016c}.
Nevertheless, reanalysis of the {\it Kepler} data, which had been baselined to remove any long-term trends
in the original analysis, showed evidence for a long-term dimming trend of $\sim 3$\% over a 4~year timescale
\citep{Montet2016}.

This secular dimming has cast doubt on the exocomet hypothesis, leading \citet{Wright2016b} to explore
alternative models for the origin of the shorter duration dimming events (i.e., the dips), including
for example the possibility of small-scale structures in the intervening
interstellar medium and artificial structures \citep{Wright2016}.
Indeed, in their ranking of the possible explanations for the dips, \citet{Wright2016b} deemed the exocomet model
less likely than an explanation involving extraterrestrial intelligence
\citep[see also][]{Abeysekara2016,Harp2016,Schuetz2016}. 
However, no strong arguments were given for the dismissal of the exocomet explanation,
except that it might be difficult to explain the secular dimming given the infrared
flux upper limits \citep[see also][]{Lacki2016}.
Others have sought to explain the secular dimming in ways not directly connected with the shorter duration dips,
such as through the accretion of a planet onto the star \citep{Metzger2016}.

In this paper we consider whether the secular dimming is indeed a serious problem for the
exocomet interpretation of the dips. 
Our underlying assumption is that both the short duration dips and the long-term secular dimming
are caused by dusty material that is unevenly distributed around a single elliptical orbit.
This is a less biased way of describing the {\it exocomet} model, since there is no requirement
that {\it exocomets} \citep[such as the Solar System-like comets of][]{Bodman2016} are involved. 
We extend the {\it scenario independent} analysis of section 4.4 of \citet{Boyajian2016},
and hence show that the secular dimming places further strong constraints on the orbit,
as well as quantify the distribution of dust around the orbit.
Note that {\it scenario independent} here means that no assumptions are made except that the short duration
dips and secular dimming are caused by material on a circumstellar orbit.
As such these constraints apply equally to natural structures (like those resulting from the disruption of
exocomets) and artificial structures (like Dyson spheres).

The layout of the paper is as follows.
In \S \ref{s:lc} we summarise the light curve, and in particular the different assumptions about secular dimming that
will be used throughout the paper.
We then use the constraints from the lack of infrared emission from circumstellar debris
to set constraints on the orbit.
This is achieved first in \S \ref{s:ir} through a heuristic model in which the dust is assumed to be
spread evenly around the orbit, and the infrared emission characterised only by its luminosity.
Then the wavelength dependence of the emission is considered, as well as issues related to the short duration dips.
Having discussed general constraints on the orbit, in \S \ref{s:dist} we use the light curve to characterise the 
uneven distribution of the dust around the orbit and consider how this
translates into the time evolution of the thermal emission, and how the timing of the infrared measurements affects
the conclusions.
Consideration of the evolution of the total infrared flux (including both dust thermal emission and starlight)
is given in \S \ref{s:stellardim}, and optical depth effects are considered in \S \ref{ss:thickedge}.
Finally in \S \ref{s:mc} we use a Monte Carlo model to consider the probability of witnessing a transit
in this system.
Conclusions are given in \S \ref{s:conc}.

\section{Light curve}
\label{s:lc}
As discussed in \S \ref{s:intro}, there remains uncertainty on the level of secular dimming that the star exhibits,
and indeed on the fraction of any observed secular dimming that originates in the same phenomenon that is causing the
shorter duration dips \citep[e.g.,][]{Makarov2016}.
Hence there is some flexibility in the observational constraints on the models presented in this paper.
Here we invoke four different interpretations of the secular dimming observations that encompass a range
of possibilities.
The resulting light curves are plotted in Fig.~\ref{fig:lczoom}.

\begin{figure*}
  \begin{center}
    \vspace{-2.7in}
    \begin{tabular}{c}
      \hspace{-2.1in}
      \includegraphics[width=3.4\columnwidth]{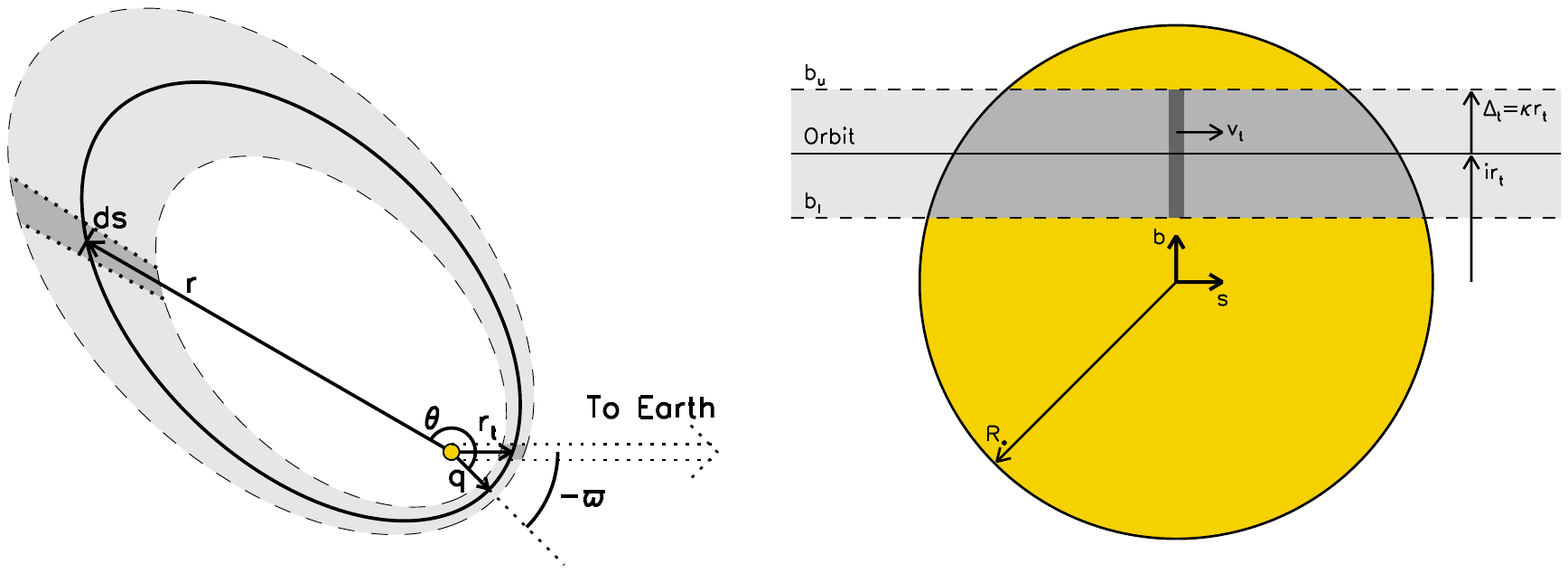}
    \end{tabular}
    \vspace{-3.1in}
    \caption{Assumed geometry of the material blocking the light from the star.
    {\bf (Left)} Face-on view showing the orientation of the pericentre $\varpi$
    and the distance of the orbit from the star at the point of transit $r_{\rm t}$.
    The size of the star and width of the distribution are exaggerated to illustrate
    the geometry.
    The width and height of the distribution are assumed to scale with distance from the star.
    The shading shows an element of the distribution at a true anomaly $\theta$ of projected
    length along the orbit of $ds$ that is discussed in \S \ref{ss:thicktemp}.
    {\bf (Right)} View along the line of sight to the star.
    The orbit crosses at an impact parameter $ir_{\rm t}$ above the centre of the
    star, but the material forms a narrow distribution about this orbit, of height and radial width
    $2\Delta_{\rm t}$ at this point, that covers the star from impact
    parameters $b_{\rm l}$ to $b_{\rm u}$.
    The shading is used to identify different
    portions of the orbit, as discussed in the text, and is the same on both left and right
    figures.
    It does not necessarily indicate the optical depth of the material at this location, which
    may vary around the orbit.
    }
   \label{fig:geom}
  \end{center}
\end{figure*}

The simplest interpretation, which is that presented in \citet{Boyajian2016} and plotted in blue on Fig.~\ref{fig:lczoom},
is that there is no secular dimming.
The next level up in terms of the amount of secular dimming, which is that presented in \citet{Montet2016} and plotted
in orange on Fig.~\ref{fig:lczoom}, considers that the only secular dimming is that seen in the {\it Kepler} data itself.
In that light curve there is no secular dimming at the start of the {\it Kepler} observations, but $\sim 3$\% dimming seen at the end.
The most extreme levels of dimming we consider again take the secular dimming seen by {\it Kepler} \citep{Montet2016}, but also
consider that the putative century-long dimming \citep{Schaefer2016} is real (and takes the form of a linear decline
of 16\% over the past century).
However, since it is unknown if the dimming persisted beyond the 1980s, we consider one interpretation in
which there is still material in front of the star at the start of the {\it Kepler} mission which is causing
a 16\% dimming level (shown in red on Fig.~\ref{fig:lczoom}),
while in another the level of dimming returned to normal shortly before {\it Kepler} started
observing (shown in green on Fig.~\ref{fig:lczoom}).

These four different light-curves will be considered in more detail in \S \ref{s:dist}, where it is further assumed
that the post-{\it Kepler} level of dimming remains constant at the level seen in
the light curve at the end of the {\it Kepler} lifetime, but the star returns to its normal brightness over the time remaining
for the assumed orbital period from when the dimming started.
For \S \ref{s:ir} the variability in the secular dimming level is not considered.
Rather, that section considers what the implications are for the thermal emission light curve if the secular dimming
level was constant, for which we use 10\% dimming as the reference level.

The short duration dips discovered by \citet{Boyajian2016} are also evident in Fig.~\ref{fig:lczoom}. 
We will make predictions for the emission from the material causing these dips in \S \ref{s:dist}.
However, these dips also provide constraints on possible orbits, with the shortest duration dip at 1568.5\,d shown
in the inset of Fig.~\ref{fig:lczoom}, which has a duration of $\sim 0.4$\,days, providing the strongest constraints.
This is discussed further in \S \ref{ss:shortdip}.

\section{Infrared constraints for dust evenly distributed around orbit}
\label{s:ir}

\subsection{Assumed geometry of the orbit}
\label{ss:geom}
The underlying assumption throughout this paper is that the material causing the short duration dimming events (i.e., the dips) and
the secular dimming shares the same orbit around a star of mass $M_\star=1.43M_\odot$ and radius $R_\star=1.58R_\odot=7.35\times10^{-3}$\,au;
other stellar parameters assumed here are a temperature $T_\star=6750$\,K, luminosity $L_\star=4.7L_\odot$, and
distance $d=392$\,pc \citep[][]{Boyajian2016,Lindegren2016}.
That orbit is defined by its orbital plane, its periastron distance $q$ and eccentricity $e$, and the orientation of its
pericentre $\varpi$ (see Fig.~\ref{fig:geom} left);
the parameters used in the paper are summarised in Table~\ref{tab:sym}.
The line of sight to the star is assumed to be very close to the orbital plane, inclined by an angle $i < R_\star/r_{\rm t}+\kappa$
to that plane (see Fig.~\ref{fig:geom} right), where $r_{\rm t}$ is the distance of the material to the star at the point of transit
and $\kappa$ is defined below.
The pericentre orientation $\varpi$ is measured relative to the mid-point of the transit, so that
\begin{equation}
  r_{\rm t} = q(1+e)/(1+e\cos{\varpi}).
  \label{eq:rt}
\end{equation}
The speed at which material crosses the star (i.e., the tangential component of its orbital velocity) is
\begin{equation}
  v_{\rm t} = h/r_{\rm t},
  \label{eq:vt}
\end{equation}
where $h=\sqrt{GM_\star q(1+e)}$ is the specific orbital angular momentum.

The assumption that all material shares exactly the same orbit cannot be exactly true, since the dip depths imply that a
significant fraction of the stellar surface is blocked (whereas a single orbit traces a narrow line across the star).
Thus a more realistic interpretation of this assumption is that the material is in a narrow distribution centred on this one orbit,
extending up to a small distance $\Delta$ both vertically and radially from the central orbit.
It is assumed that the height and width of this
distribution scales linearly with distance from the star so that $\Delta=\kappa r$ for some constant $\kappa$.
Thus at the point of transit, the material is seen as a horizontal band covering impact parameters above the
centre of the star from $b_{\rm l} = (i-\kappa)r_{\rm t}$ to $b_{\rm u} = (i+\kappa)r_{\rm t}$
(see Fig.~\ref{fig:geom} right).

This means that the fraction of the stellar disk covered by the material
(which is the area of the medium and dark shaded regions in Fig.~\ref{fig:geom} right
divided by $\pi R_\star^2$) is given by
\begin{eqnarray}
  \Omega_\star & = & \pi^{-1}[\cos{^{-1}(b_{\rm l}/R_\star)}
                              - (b_{\rm l}/R_\star) \sqrt{1-(b_{\rm l}/R_\star)^2} \nonumber \\
               &   &          - \cos{^{-1}(b_{\rm u}/R_\star)}
                              + (b_{\rm u}/R_\star) \sqrt{1-(b_{\rm u}/R_\star)^2}],
  \label{eq:omstar}
\end{eqnarray}
where for this calculation it should be assumed that $b_{\rm u} = R_\star$ or $b_{\rm l} = -R_\star$ if  
either of these impact parameters falls outside the star.
For distributions that are much narrower than the star when seen in projection, i.e. those with
$\kappa \ll R_\star/r_{\rm t}$, this results in
\begin{equation}
  \Omega_\star \approx (4/\pi)(r_{\rm t}/R_\star)\kappa\sqrt{1-(ir_{\rm t}/R_\star)^2},
  \label{eq:omst}
\end{equation}
which is the ratio of the height of the band times the length of the orbit projected onto the
star divided by the area of the star.

The covering fraction $\Omega_\star$ is related to the fraction of starlight that is removed from our line of sight by the
material, which we call the dip depth or dimming $\delta_\lambda$, by the optical depth of the material due to extinction
$\tau_{\rm ext}$, noting that both of these quantities are dependent on wavelength $\lambda$.
When considering secular dimming we will assume that at any given time the optical depth is uniform across the face
of the star (e.g., see the medium shading in Fig.~\ref{fig:geom} right).
Here the star is also assumed to be of uniform brightness (i.e., ignoring limb darkening for example)
so that 
\begin{equation}
  \delta_\lambda = [1-\exp{(-\tau_{\rm ext})}]\Omega_\star.
  \label{eq:deltadef}
\end{equation}
This means that the maximum dip depth or dimming is $\Omega_\star$, and so for a given level of
observed dimming $\delta_\lambda$, there is already a lower limit on the height of the distribution
such that $\kappa>\kappa_{\rm min}$, where
\begin{equation}
  \kappa_{\rm min} = \delta_\lambda (\pi/4) R_\star/r_{\rm t},
  \label{eq:kmin}
\end{equation}
with this limit only achieved for the case of an infinite optical depth and an orbit exactly aligned with
the line of sight. 

When considering the short duration dips in the light curve we will assume that the optical depth 
toward the stellar disk is independent of impact parameter, but can vary significantly along the orbit, or rather
with azimuthal distance as seen toward the star (e.g., see the darker shading in Fig.~\ref{fig:geom} right).
A more general assumption would have an optical depth which also varied with impact parameter.

\subsection{Assumed dust properties}
\label{ss:dust}
Throughout this paper we will make one of two assumptions about the properties of the dust orbiting KIC8462852.
To simplify the analytics, and so gain insight into the problem, we first assume that the dust behaves like a black body.
However, for a more realistic prescription we later assume that the dust composition is that of astronomical silicate
grains that have a density of 3.3\,g\,cm$^{-3}$ \citep[][]{Laor1993};
Mie theory is used to calculate the dust optical properties, i.e., the absorption and extinction coefficients
at wavelength $\lambda$ for a grain of different diameter $D$, $Q_{\rm abs}(\lambda,D)$ and $Q_{\rm ext}(\lambda,D)$,
respectively.
We also assume a size distribution in which the fraction of the total cross-sectional area $\sigma_{\rm tot}$ that is in
particles with diameters $D$ to $D+dD$ is given by $\sigma(D)dD$ where $\sigma(D) \propto D^{2-\alpha}$ and $\alpha=7/2$.
This size distribution is assumed to extend from a size large enough to be inconsequential (since most of the cross-sectional
area is in small grains for $\alpha>3$), down to a minimum size set by the radiation pressure blow-out limit.
Since the orbit of the parent body of the dust is eccentric, the blow-out limit would be different for dust released
at different points around the orbit \citep{Burns1979,Wyatt2010,Lohne2017}.
For simplicity we avoid such considerations and simply use for the minimum size that of the blow-out limit for a circular orbit
(i.e., that for which the ratio of the radiation force to that of the star's gravity
$\beta=0.5$), which is for particles with diameter $2.3$\,$\mu$m for the dust and stellar properties
assumed here.

\begin{figure}
  \begin{center}
    \begin{tabular}{c}
      \includegraphics[width=1\columnwidth]{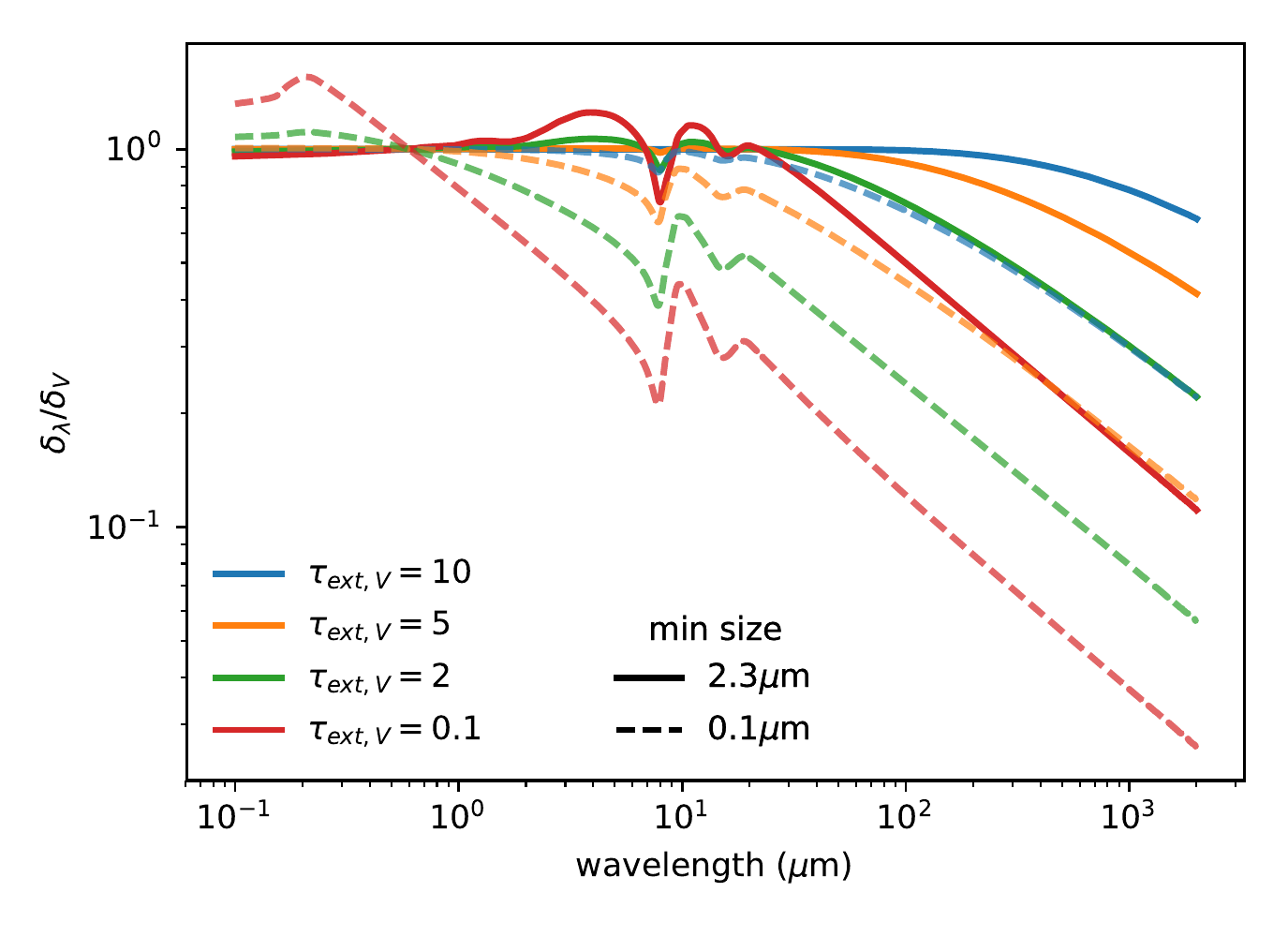}
    \end{tabular}
    \caption{The ratio of dimming expected at a wavelength $\lambda$ to that in the optical,
    $\delta_\lambda / \delta_{\rm V}$, as a function of wavelength for the dust optical properties
    and size distribution of \S \ref{ss:dust} (solid lines).
    This ratio also depends on optical depth, and so the different coloured lines show different levels of V-band
    extinction $\tau_{\rm ext,V}=0.1,2,5,10$ (red, green, orange and blue, respectively).
    To highlight the sensitivity to the assumed dust properties, the dashed lines show the ratio for the
    same assumptions except with a minimum grain size of 0.1\,$\mu$m.
    }
   \label{fig:dldv}
  \end{center}
\end{figure}

The above dust properties come into the following analysis through two factors. 
One of these is the absorption coefficient of the dust
$Q_{\rm abs}$ averaged over the size distribution $\sigma(D)$ and stellar spectrum $F_{\nu \star}(\lambda)$,
\begin{equation}
  \langle Q_{\rm abs} \rangle_{D,\star} =
     \frac{\int F_{\nu \star}(\lambda) \int \sigma(D) Q_{\rm abs}(\lambda,D) dD d\lambda}
                                    {\int F_{\nu \star}(\lambda) d\lambda \sigma_{\rm tot}},
  \label{eq:qabs}
\end{equation}
which for the above particle properties is $0.9$.
The other factor is the extinction coefficient of the dust $Q_{\rm ext}$ averaged over the size distribution
at the wavelength $\lambda$ at which dimming is measured,
\begin{equation}
  \langle Q_{\rm ext} \rangle_{D} = \int (\sigma(D)/\sigma_{\rm tot}) Q_{\rm ext}(\lambda,D) dD,
  \label{eq:qext}
\end{equation}
noting that this quantity is wavelength dependent.
For the above particle properties $\langle Q_{\rm ext} \rangle_{D} = 2.1$ for the Kepler bandpass,
and has a similar value for all wavelengths below $\sim 3$\,$\mu$m
(since $\delta_\lambda \propto \langle Q_{\rm ext} \rangle_{D}$ for optically thin distributions
and so the wavelength dependence of $\langle Q_{\rm ext} \rangle_{D}$ can be inferred from
the solid red line in Fig.~\ref{fig:dldv}).
Note that this treatment does not include light that is scattered into the line of sight,
the effect of which on the light curve is typically much smaller than the effect of extinction
\citep[for cometary dust at a distance of 1\,au from the star, the difference is about
one order of magnitude;][]{Lecavelier1999}.
Another motivation for not including this component is the lack of strong positive bumps in the 
\textit{Kepler} light curve.
We refer to \citet{Lamers1997} for a description of how scattering effects could be modelled.

Given that it is possible to measure dimming at multiple wavelengths it is worth noting that our assumptions
result in an extinction coefficient that is roughly independent of wavelength at optical wavelengths,
only falling off in the infrared.
For example, Fig.~\ref{fig:dldv} shows the ratio of $\delta_\lambda/\delta_{\rm V}$, which for optically
thin distributions is equivalent to
$\langle Q_{{\rm ext},\lambda} \rangle_{\rm D} / \langle Q_{\rm ext,V} \rangle_{\rm D}$.
Thus the assumed dust properties are consistent with the grey extinction inferred by \citet{Meng2017}.
This is primarily because the minimum grain size in the assumed distribution is larger than the wavelength
of observation.
This minimum size was assumed to be set by radiation pressure, but we find that wavelength dependent
extinction with $Q_{\rm ext} \propto \lambda^{-0.5}$ in the optical would be expected if the distribution
had been assumed to extend down to grain sizes smaller than $0.1$\,$\mu$m (see Fig.~\ref{fig:dldv}).
Fig.~\ref{fig:dldv} also shows that the grey extinction inferred by Meng et al. (submitted) would
be expected if the distribution is optically thick, even for very small minimum grain sizes.
Note that the secular dimming could have a different wavelength dependence to the dips if the
two features have different dust size distributions;
e.g., in an exocomet model in which dust is released from planetesimal-sized bodies, the dips could
contain small dust that is quickly expelled by radiation pressure, while the dimming could originate
in the larger dust that remains gravitationally bound.

It is also important to note that in most of what follows we have at first implicitly assumed that
the contribution of stellar emission has been perfectly subtracted from any observed infrared fluxes,
so that the only emission that is seen is that from the thermal emission of the dust.
This subtraction will always be imperfect because of uncertainties in the absolute level of photospheric
emission (typically at the level of $\sim 1$\%).
However, this subtraction is further complicated because the wavelength dependence of the dimming
is not known a priori.
Nevertheless, the resulting decrease in brightness can be predicted for an assumed dust composition and
size distribution from Fig.~\ref{fig:dldv}, and this is considered further in \S \ref{s:stellardim}.

\subsection{Fractional luminosity for evenly distributed dust}
\label{ss:flim}
In this section we will assume that the total cross-sectional area of material $\sigma_{\rm tot}$ is spread evenly around the
orbit; i.e., that the line density of material is inversely proportional to its orbital velocity.

\subsubsection{Fractional luminosity calculation}
\label{sss:flum}
For an even optically thin distribution of black body dust, equation~46 of \citet{Wyatt2010} shows that
the fractional luminosity of the material (i.e., the ratio of the infrared luminosity from the
material to that of the star, $f=L_{\rm IR}/L_\star$) is given by 
\begin{equation}
  f_{\rm BB} = \sigma_{\rm tot}\left(4 \pi q^2(1-e)^{-2}\sqrt{1-e^2}\right)^{-1}.
  \label{eq:f}
\end{equation}

In so far as equation~\ref{eq:f} just considers the fraction of starlight intercepted by the material, it is independent of
the material properties.
However, if the material is inefficient at absorbing the starlight (e.g., because it is made up of particles smaller than the
wavelength of the starlight) then the fractional luminosity would be lower.
Indeed, equation~\ref{eq:f} should include an additional factor of $\langle Q_{\rm abs} \rangle_{D,\star}$ 
(eq.~\ref{eq:qabs}) on the right hand side. 
This factor is shown below to be relatively unimportant for the following calculations, since material that
absorbs less causes both less infrared emission and less dimming.
Nevertheless, the material properties will be important later when we wish to consider the wavelength at which
the material re-emits the absorbed energy (which is considered in \S \ref{ss:irobs}).

One assumption in the calculation of eq.~\ref{eq:f} is that the dust distribution is optically thin along the line of sight
to the star (so that all of the cross-sectional area is available to absorb starlight). 
Clearly this cannot be the case if the total cross-sectional area is too large.
For the geometry specified in \S \ref{ss:geom}, the radial optical depth is independent of longitude (i.e.,
independent of whether we are considering the portion of the orbit near pericentre, apocentre, or in between, since the
distribution is vertically broader further from the star).
Starlight at wavelength $\lambda$ is attenuated by a factor of $\exp{(-\tau_{\rm ext})}$ as it passes through the ring.
To acknowledge that some fraction of the extinction is due to light scattered out of the beam we define
$\exp{(-\tau_{\rm abs})}$ to be the attenuation due to dust absorption, so that the optical depth can be accounted for in
eq.~\ref{eq:f} by an additional factor of $[1-\exp{(-\langle\tau_{\rm abs}\rangle_\star)}]/\langle\tau_{\rm abs}\rangle_\star$
on the right hand side, where the angle brackets indicate that the optical depth has been averaged over the stellar spectrum.
This factor is again found to be unimportant for the calculation in \S \ref{sss:dfrel} of the resulting
dimming, because if the cross-sectional area contributing to the fractional luminosity is reduced due to optical
depth effects, then so is the level of dimming.

Taking account of the additional factors from the last two paragraphs, the fractional luminosity is thus
\begin{equation}
  f = f_{\rm BB} \langle Q_{\rm abs} \rangle_{D,\star} [1-\exp{(-\langle\tau_{\rm abs}\rangle_\star)}]/\langle\tau_{\rm abs}\rangle_\star.
  \label{eq:f2}
\end{equation}

Given the definition of the optical depth, an alternative way of deriving the fractional luminosity is
\begin{equation}
  f=\kappa[1-\exp{(-\langle\tau_{\rm abs}\rangle_\star)}].
  \label{eq:fnew}
\end{equation}
Equation~\ref{eq:fnew} shows that the maximum possible fractional luminosity for the assumed geometry
is $f_{\rm max}=\kappa$, which cannot be exceeded because at this point the material is radially
optically thick and so adding more cross-sectional area does not increase the amount
of starlight absorbed.
For optically thin distributions eq.~\ref{eq:fnew} can be simplified to
$f=\kappa \langle\tau_{\rm abs}\rangle_\star$.

Note that the derivation above has only accounted for the optical depth along the line of sight from the star to the dust.
That optical depth affects the amount of light absorbed by the dust and so the thermal luminosity of the dust (since the
absorbed energy must be reemitted).
However, the derivation does not account for the optical depth along our line of sight to the dust.
If this optical depth is large it could mean that the inferred luminosity is different when observed in different directions,
notably with a lower luminosity inferred for edge-on orientations such as that considered here (see \S \ref{ss:thickedge}).

\subsubsection{Optical depth}
\label{sss:tau}
Combining eq.~\ref{eq:fnew} with eq.~\ref{eq:f2} shows that the optical depth is given by
\begin{equation}
  \langle\tau_{\rm abs}\rangle_\star = \langle Q_{\rm abs} \rangle_{D,\star} f_{\rm BB} / \kappa.
  \label{eq:taustar2}
\end{equation}
This can be derived another way which provides further insight into the origin of this expression.

First consider the plane that includes both the normal to the orbital plane and the line of sight
to the point in the orbit at the mid-point of the transit (which is approximately the line of sight to the star).
The amount of area crossing that plane per unit time must be constant if material is evenly distributed
around the orbit, and since $\sigma_{\rm tot}$ must pass by in one orbital period ($t_{\rm per}$) this means that
cross-sectional area passes through at a rate $\dot{\sigma} = \sigma_{\rm tot} / t_{\rm per}$.
Given equation~\ref{eq:f}, this means that
\begin{equation}
  \dot{\sigma} = 2hf_{\rm BB},
  \label{eq:dsigmadt}
\end{equation}
remembering that $h$ is the specific orbital angular momentum.
Importantly, since the orientation of this plane relative to the pericentre direction $\varpi$ was not defined,
this is true for any line of sight (i.e., whether that line of sight is in the direction of the pericentre, apocentre,
or in between).

The amount of material in front of the star at any given time can be determined by dividing
equation~\ref{eq:dsigmadt} by the transverse velocity of the material at this point $v_{\rm t}$ (eq.~\ref{eq:vt}).
This gives the projected line density of material, which when multiplied by the length of the orbit projected onto the
star gives the area of dust in front of the star for narrow dust distributions (i.e., those with $\kappa \ll R_\star/r_{\rm t}$).
More generally this area is $(\pi R_\star^2 \Omega_\star)f_{\rm BB}/\kappa$.
The factor in brackets is the area of the stellar disk covered by material (i.e., the area of the medium plus dark shaded region in
Fig.~\ref{fig:geom} right), and so the optical depth is simply $f_{\rm BB}/\kappa$ times a factor which accounts
for the efficiency with which starlight is absorbed by the dust, which results in eq.~\ref{eq:taustar2}.
It is thus easy to determine that the extinction optical depth is given by
\begin{equation}
  \tau_{\rm ext} = \langle Q_{\rm ext} \rangle_{D} f_{\rm BB} / \kappa.
  \label{eq:tauext}
\end{equation}
Thus the ratio of the extinction to absorption optical depths is 2.4 for the particle properties
assumed in \S \ref{ss:dust}.

\subsubsection{Dimming to fractional luminosity relation}
\label{sss:dfrel}
Here we show that there is a simple relation between the fractional luminosity and the quantity
of material blocking our line of sight to the star.
This is most readily achieved by combining eq.~\ref{eq:fnew} with eq.~\ref{eq:deltadef} to find that
\begin{equation}
  f= \delta_\lambda \left( \frac{\kappa}{\Omega_\star} \right)
                    \left[ \frac{ 1-\exp{ (-\langle \tau_{\rm abs} \rangle_\star) } }
                                { 1-\exp{ (-\tau_{\rm ext})                         } } \right].
  \label{eq:delta2}
\end{equation}

The factor in square brackets in eq.~\ref{eq:delta2} accounts for the fact that dimming is extinction measured
at a single wavelength, whereas disk luminosity involves an integral of absorption over the stellar spectrum.
However, assuming that the optical depth is not significantly different across the stellar spectrum
and that the dust absorption is similar to its extinction, this factor is of order unity.
That is, the dependence on optical depth drops out of equation~\ref{eq:delta2}, because this affects dimming
and luminosity (roughly) equally;
e.g., for the assumed dust properties this factor is unity for optically thick distributions and $\sim 0.4$
for optically thin distributions.
Similarly, if the dust blocks the starlight inefficiently, this does not significantly affect the calculation in
eq.~\ref{eq:delta2}, since this would affect both the fractional luminosity and the fraction of light blocked (roughly)
equally.

\begin{figure}
  \begin{center}
    \vspace{-0.8in}
    \begin{tabular}{c}
      \hspace{-0.9in}
      \includegraphics[width=1.4\columnwidth]{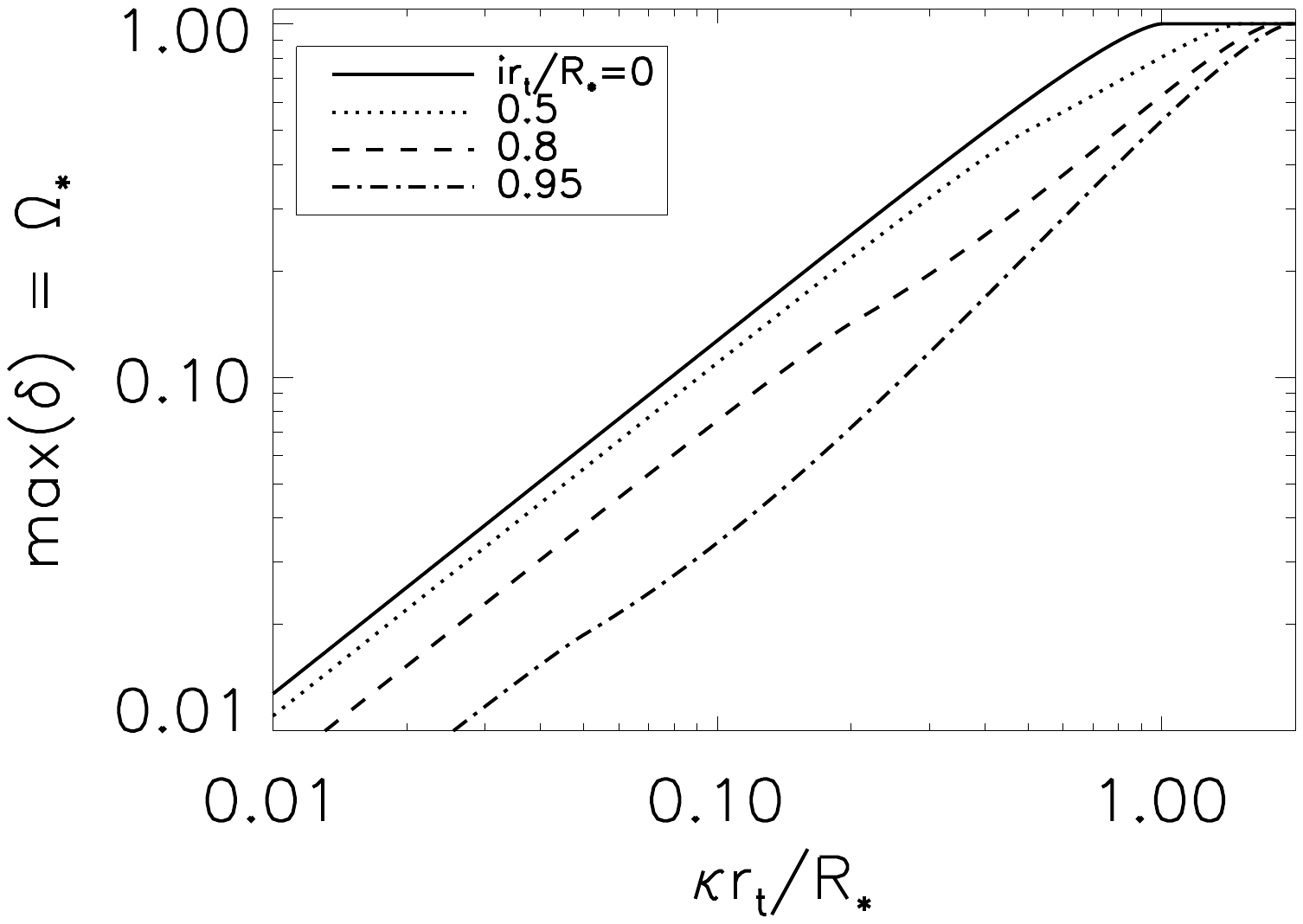} \\[-3.5in]
      \hspace{-0.9in}
      \includegraphics[width=1.4\columnwidth]{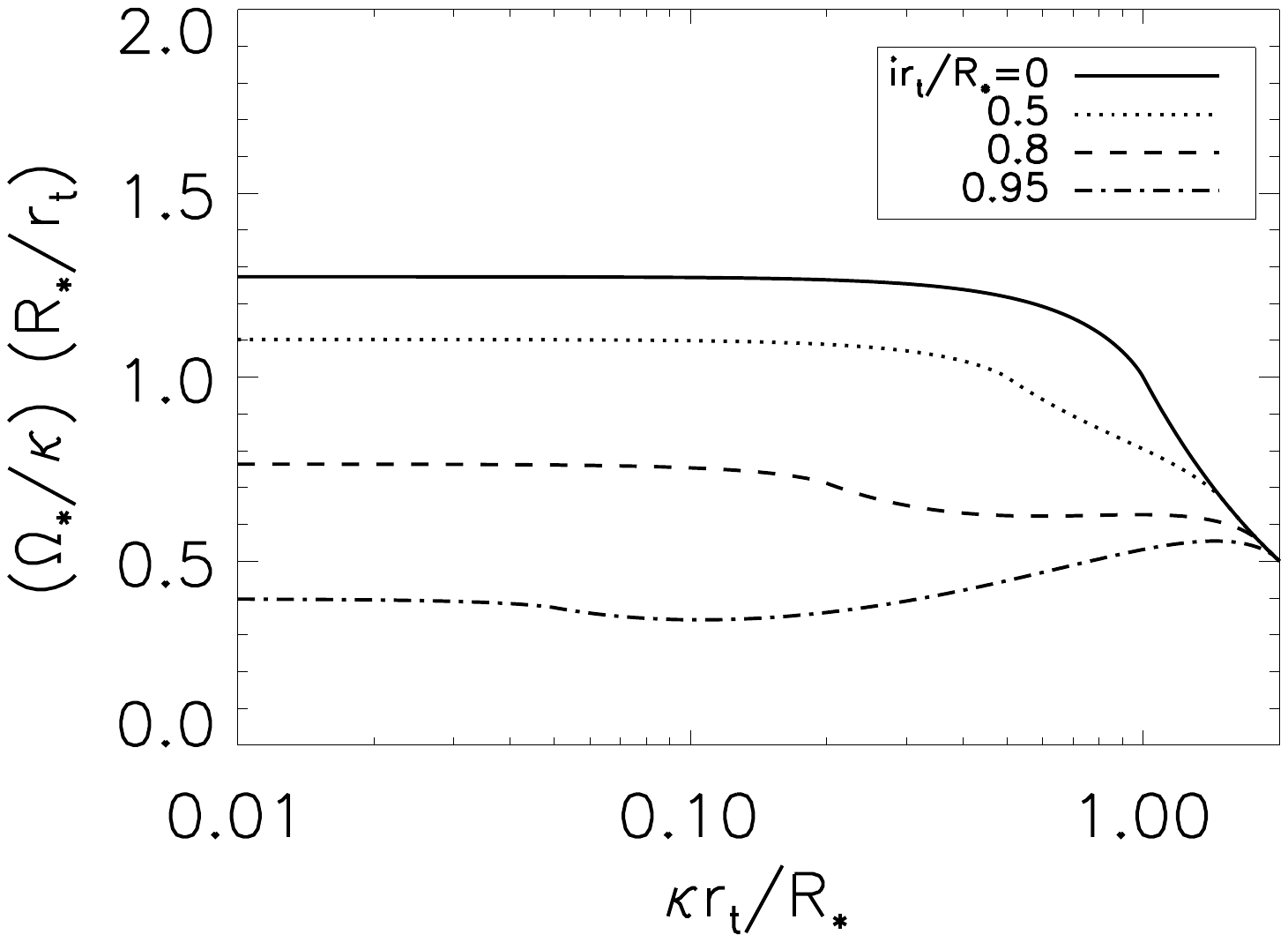} \\[-0.6in]
    \end{tabular}
    \vspace{-2.3in}
    \caption{Geometrical quantities that translate into observables.
    {\bf (Top)} The fraction of the star covered by material, $\Omega_\star$, as a function of
    the height of the distribution, $\kappa$, and the inclination of the orbit, $i$.
    This covering fraction defines the maximum possible level of dimming $\delta$ for the given geometry.
    {\bf (Bottom)} For dust that is evenly distributed around the orbit, the ratio of dimming to
    fractional luminosity is approximately given by $\Omega_\star/\kappa$ (eq.~\ref{eq:delta2}).
    Since this plot shows $(\Omega_\star/\kappa)(R_\star/r_{\rm t})$ as a function of $\kappa$ and $i$, 
    the ratio of dimming to fractional luminosity is given by the y-axis multiplied by
    $(r_{\rm t}/R_\star)$.
    Combined with the top plot, this shows that for $>10$\% dimming, the fractional luminosity is
    approximately a factor $R_\star/r_{\rm t}$ smaller than the dimming.
    }
    \label{fig:geom2}
  \end{center}
\end{figure}

If the factor in square brackets in eq.~\ref{eq:delta2} is unity, this means that the ratio of the dimming to fractional luminosity is
just given by Fig.~\ref{fig:geom2} bottom.
For dimming of $>10\%$ the top panel of Fig.~\ref{fig:geom2} shows that there is a minimum height for the distribution
which means that the factor plotted in the bottom panel of Fig.~\ref{fig:geom2} is of order unity.
This means that the fractional luminosity is a factor of approximately $R_\star/r_{\rm t}$ smaller than the dimming;
this ratio is only dependent on the distance at which material transits in front
of the star, and is independent of whether that transit is at pericentre, apocentre or inbetween.

The implication of equation~\ref{eq:delta2} is immediately apparent by considering Figs.~8 and 10 from
\citet{Boyajian2016}, where it was shown that the infrared constraints require the thermal emission
to have a fractional luminosity less than $\rm{max}(f_{\rm obs}) \approx 4 \times 10^{-4}$.
This means that, if the secular dimming is such that of order ${\rm min}(\delta_{\rm obs})=10$\% of the starlight is continuously
blocked, this requires the material to be passing at least ${\rm min}(r_{\rm t})$ from the star, where
\begin{equation}
  {\rm min}(r_{\rm t})/R_\star = (\pi/4){\rm min}(\delta_{\rm obs})/\rm{max}(f_{\rm obs}) \approx 200,
  \label{eq:rmin}
\end{equation}
and we have used the assumption that the distribution is much narrower than the star when seen in projection
to get the factor $\pi/4$ (see eq.~\ref{eq:omst}).
Thus, for $R_\star=1.58R_\odot$, the material must be passing at least $1.5$\,au from the star.
Again, note that this minimum distance is independent of the orientation, meaning that as long as this constraint
is met, and as long as the infrared constraints really do result in a maximum fractional luminosity that is independent
of the orbit (which is not quite true as discussed in \S \ref{ss:irobs}), we could be observing the material
pass through either pericentre or apocentre at this distance.

\subsection{Observed IR constraints for even dust distribution}
\label{ss:irobs}
\begin{figure}
  \begin{center}
    \begin{tabular}{c}
      \includegraphics[width=1\columnwidth]{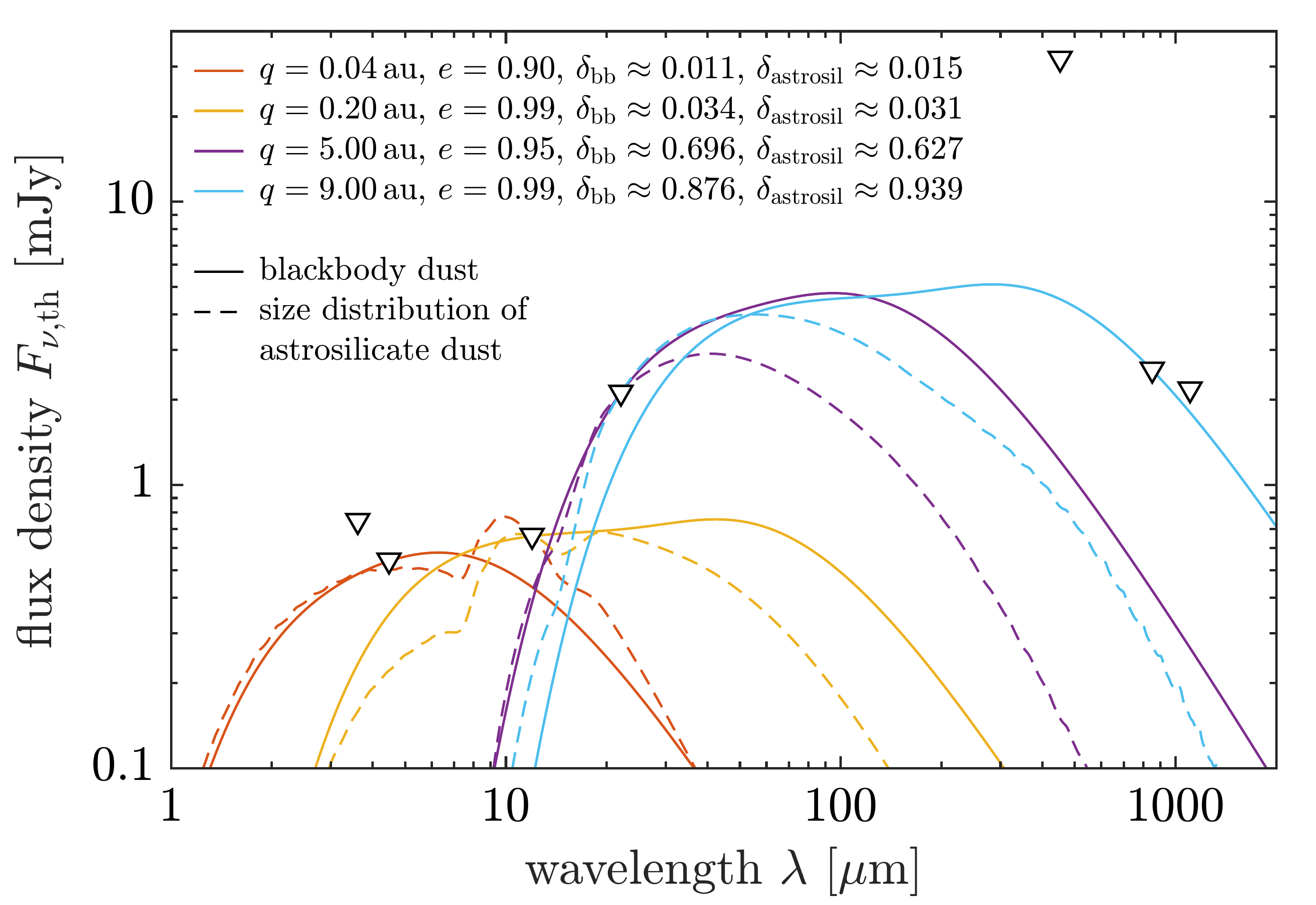}
    \end{tabular}
    \caption{Spectral energy distribution of KIC8462852.
    The downward pointing triangles give the upper limits to the dust emission from Table~\ref{tab:fir}.
    The different coloured solid lines show the emission spectrum for optically thin black body dust with
    orbital parameters given in the legend that has had the dust level increased until one of
    the observed upper limits is reached.
    The dashed lines show the spectra for the same orbital properties, but assuming
    the grain composition and size distribution given in \S \ref{ss:dust}.
    The resulting dimming level $\delta$ for the two cases is given in the legend assuming
    the transit occurs at pericentre ($\varpi=0$) with an impact parameter $b=0$.
    }
   \label{fig:spec}
  \end{center}
\end{figure}

The assumption of \S \ref{ss:flim} was that the infrared observations resulted in a uniform constraint on the
fractional luminosity of the dust distribution, which is not quite true.
In Table~\ref{tab:fir} we give the upper limits on the disk flux that will be used in this paper, and these
are plotted in Fig.~\ref{fig:spec}.
In \S \ref{s:dist} we will acknowledge that these limits were measured at a specific epoch, but for this section we will
consider that these limits apply at all times.

\begin{table*}
  \centering
  \caption{Observed upper limits ($3\sigma$) on excess thermal emission from KIC8462852 as a function
  of wavelength used to create Figs.~\ref{fig:spec} and \ref{fig:ir1}.
  Observations at 3.6 and 4.5\,$\mu$m were made with {\it Spitzer}, and those at 3.4 and 4.6\,$\mu$m with {\it WISE}. 
  The date of the observation becomes relevant in \S \ref{s:dist}, and is shown on Fig.~\ref{fig:irvt},
  but for \S \ref{s:ir} the infrared emission is assumed to be constant and the analysis
  uses the lowest limit in the table for each waveband. }
  \label{tab:fir}
  \begin{tabular}{llll}
     \hline
     Wavelength              & Date             & Disk flux             & Reference  \\
     \hline
     3.4\,$\mu$m             & 2010 May 14, 2010 Nov 9, 2014 May 16, 2014 Nov 13, 2015 May 15, 2015 Nov 7   & $<2$\,mJy  & This paper ({\it ALLWISE}) \\
     3.6\,$\mu$m             & 2015 Jan 18      & $<0.75$\,mJy          & \citet{Marengo2015} \\
     4.5\,$\mu$m             & 2015 Jan 18      & $<0.54$\,mJy          & \citet{Marengo2015} \\
     4.6\,$\mu$m             & 2010 May 14, 2010 Nov 9, 2014 May 16, 2014 Nov 13, 2015 May 15, 2015 Nov 7   & $<1$\,mJy  & This paper ({\it ALLWISE}) \\
     12\,$\mu$m              & 2010 May 14      & $<0.66$\,mJy          & This paper ({\it ALLWISE}) \\
     22\,$\mu$m              & 2010 May 14      & $<2.11$\,mJy          & This paper ({\it ALLWISE}) \\
     450\,$\mu$m             & 2015 Oct 26-29   & $<32.1$\,mJy          & \citet{Thompson2016} \\
     850\,$\mu$m             & 2015 Oct 26-29   & $<2.55$\,mJy          & \citet{Thompson2016} \\
     1100\,$\mu$m            & 2015 Nov 10      & $<2.19$\,mJy          & \citet{Thompson2016} \\
     \hline
  \end{tabular}
\end{table*}

It was prior versions of these limits which were used to make Fig.~8 of \citet{Boyajian2016} to get the limits
on the dust fractional luminosity under the assumption that the emission could be approximated by a
single temperature black body. 
In Fig.~\ref{fig:spec} we assume that the dust is uniformly distributed around a given orbit and determine the
resulting emission spectrum.
For the first calculation (solid lines) the dust is assumed to have the black body temperature given its distance from
a star with parameters given in \S \ref{ss:geom}.
This calculation was then repeated (dashed lines) using the realistic grain properties and size distribution described in
\S \ref{ss:dust}.
The method for calculating the spectrum of the thermal emission is equivalent to that given in
eqs.~37-39 of \citet{Wyatt2010}:
\begin{eqnarray}
  F_{\nu,{\rm th}} & = & 2.35 \times 10^{-11} d^{-2}
              \int_0^{2\pi} (d\sigma/d\theta) \int_{D_{\rm{min}}}^{D_{\rm{max}}} Q_{\rm{abs}}(\lambda,D) \times \nonumber \\
        &   &
                 B_\nu(\lambda,T[D,r(\theta)])
                 \bar{\sigma}(D) dD d\theta, \label{eq:fnu} \\
  d\sigma/d\theta & = & (\sigma_{\rm tot} / 2\pi)(1-e^2)^{3/2}(1+e\cos{\theta})^{-2},  \label{eq:dsdth} \\
  T(D,r)  & = &  \left[ \frac{\langle Q_{\rm{abs}}(D,\lambda) \rangle_{T_\star}}
                   {\langle Q_{\rm{abs}}(D,\lambda) \rangle_{T(D,r)}} \right]^{1/4}
                   T_{\rm{bb}}(r), \label{eq:tdr} \\
  T_{\rm{bb}}(r) & = & 278.3 L_\star^{1/4} r^{-1/2}, \label{eq:tbb}
\end{eqnarray}
where $F_{\nu,{\rm th}}$ is in Jy, $d$ is distance in pc,
$d\sigma/d\theta$ is the total cross-sectional area per longitude in au$^{2}$ per radian,
$B_\nu$ is in Jy\,sr$^{-1}$,
$\bar{\sigma}(D)dD$ is the fraction of the cross-sectional area in sizes $D$ to $D+dD$,
$\langle Q_{\rm{abs}} \rangle_T$ means $Q_{\rm{abs}}$ averaged over a
black body spectrum of temperature $T$, $L_\star$ is in $L_\odot$, and
$r$ is distance from the star in au.
These equations implicitly assume that the emission is optically thin.
A simple way of including optical depth effects which is used in this section is to add
a factor $[1-\exp{(-\langle\tau_{\rm abs}\rangle_\star)}]/\langle\tau_{\rm abs}\rangle_\star$ to
the right hand side of eq.~\ref{eq:fnu}.
This assumes that the emission is isotropic and that the dust emits at the
same temperature as it would in the optically thin regime (which is calculated in eq.~\ref{eq:tdr}).
A more detailed prescription for optical depth effects is given in \S \ref{ss:thickedge}.

Considering first the lines for black body dust on Fig.~\ref{fig:spec}, it is evident that the spectrum resembles
the superposition of two black bodies, a hot component associated with material near pericentre, and a cold
component associated with material near apocentre.
Comparing the two lines for the different dust assumptions on Fig.~\ref{fig:spec} shows that there are three
main consequences of using realistic grains:
the emission spectrum is in general hotter (because the small grains that dominate the cross-sectional area
are heated above the black body temperature),
the emission falls off faster at longer wavelengths (because the same small grains emit inefficiently at long
wavelengths),
and some spectral features appear as a result of the composition.
Since the choice of dust composition was somewhat arbitrary, we are not concerned about the details.
However, the first two points apply to all compositions and so there will be a systematic difference
in the constraints we derive between black body and realistic grains.

\begin{figure*}
  \begin{center}
    \begin{tabular}{cc}
      \includegraphics[width=0.95\columnwidth]{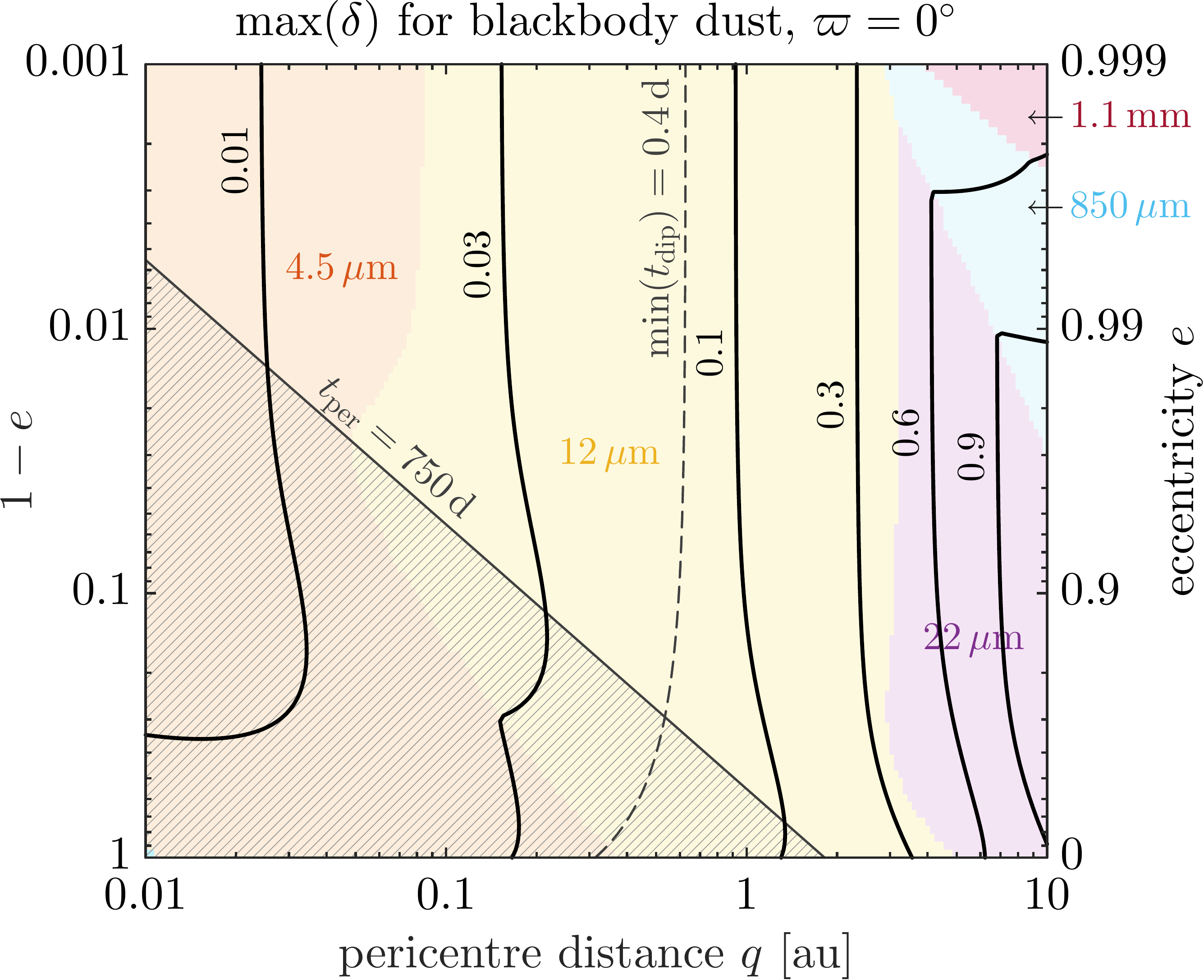} &
      \includegraphics[width=0.95\columnwidth]{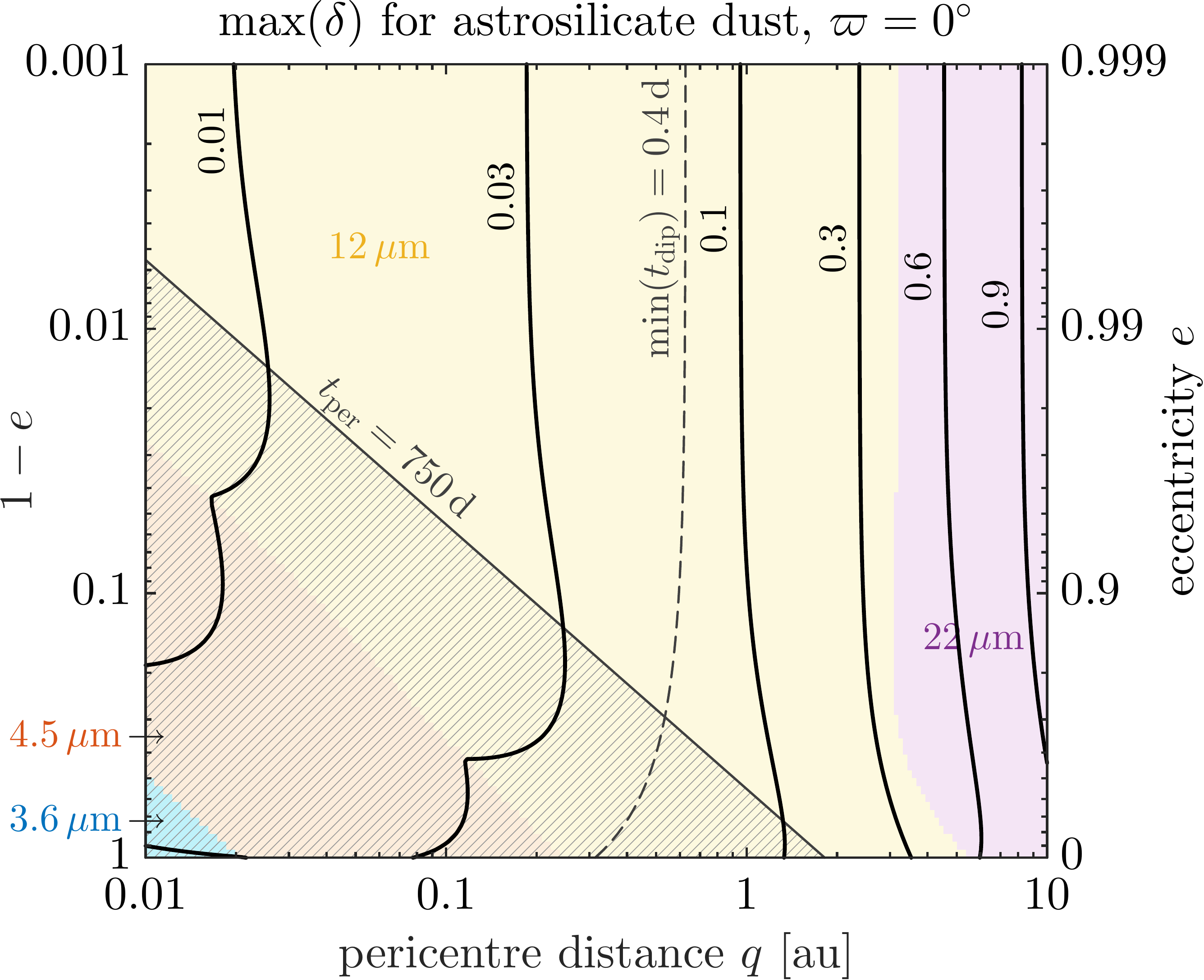} \\[0.15in]
      \includegraphics[width=0.95\columnwidth]{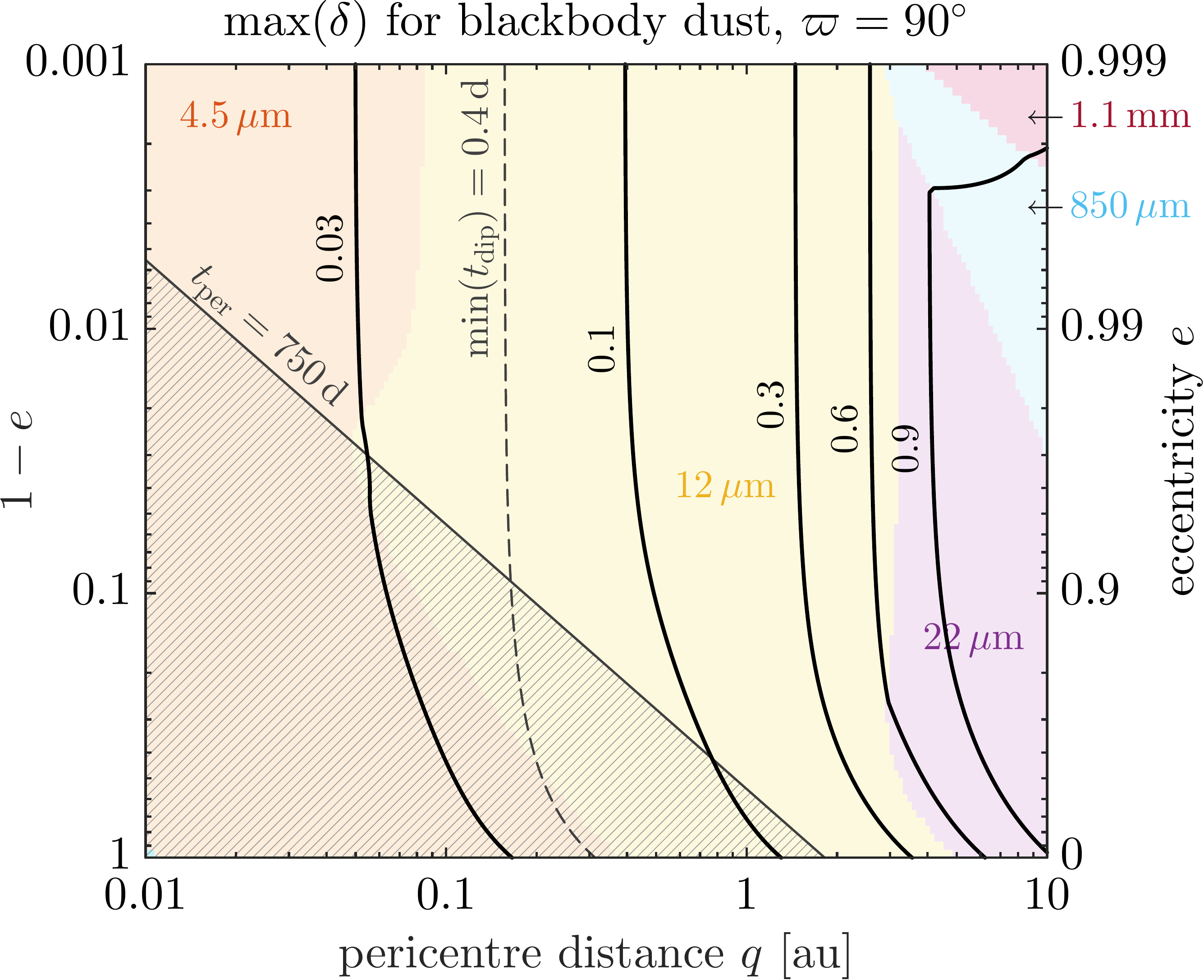} &
      \includegraphics[width=0.95\columnwidth]{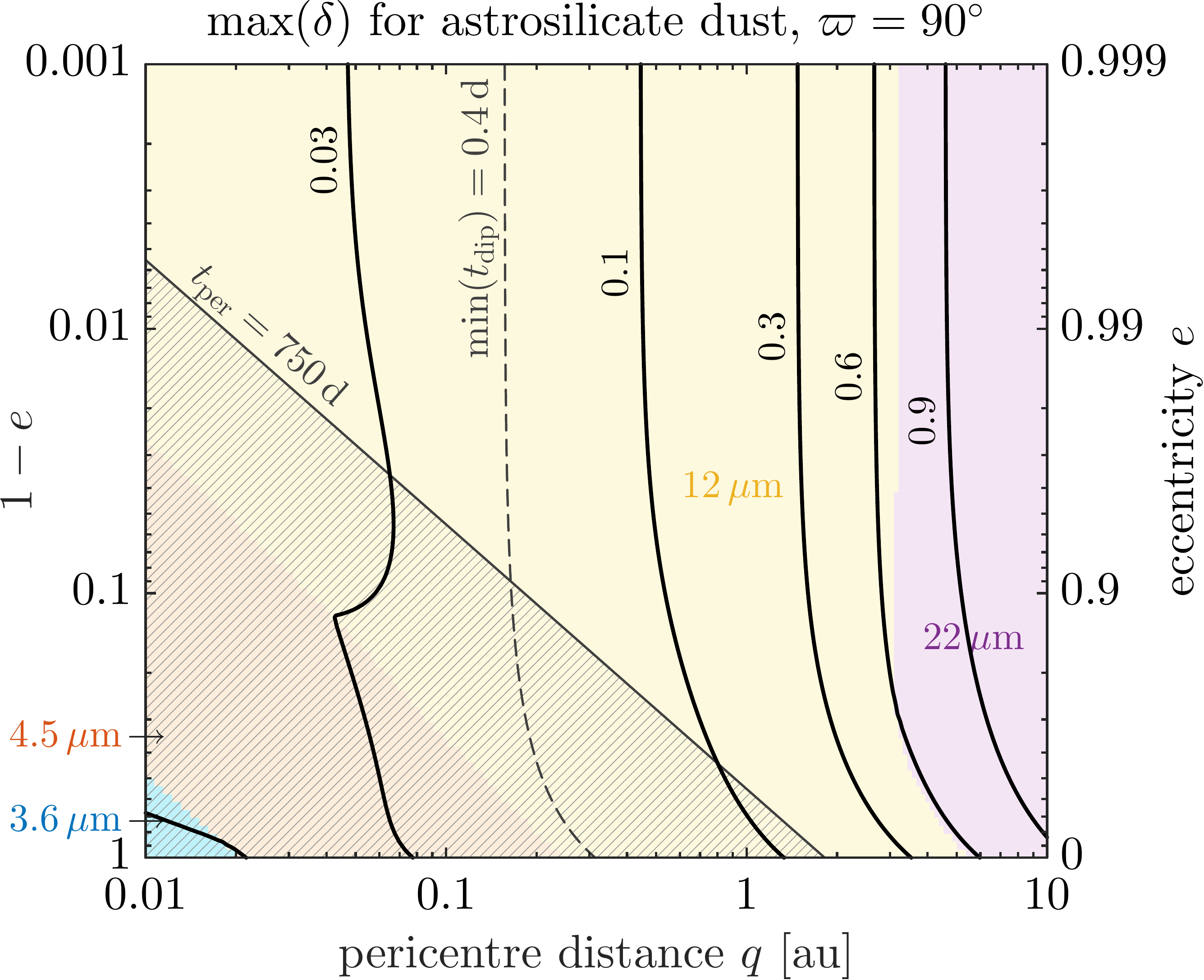} \\[0.15in]
      \includegraphics[width=0.95\columnwidth]{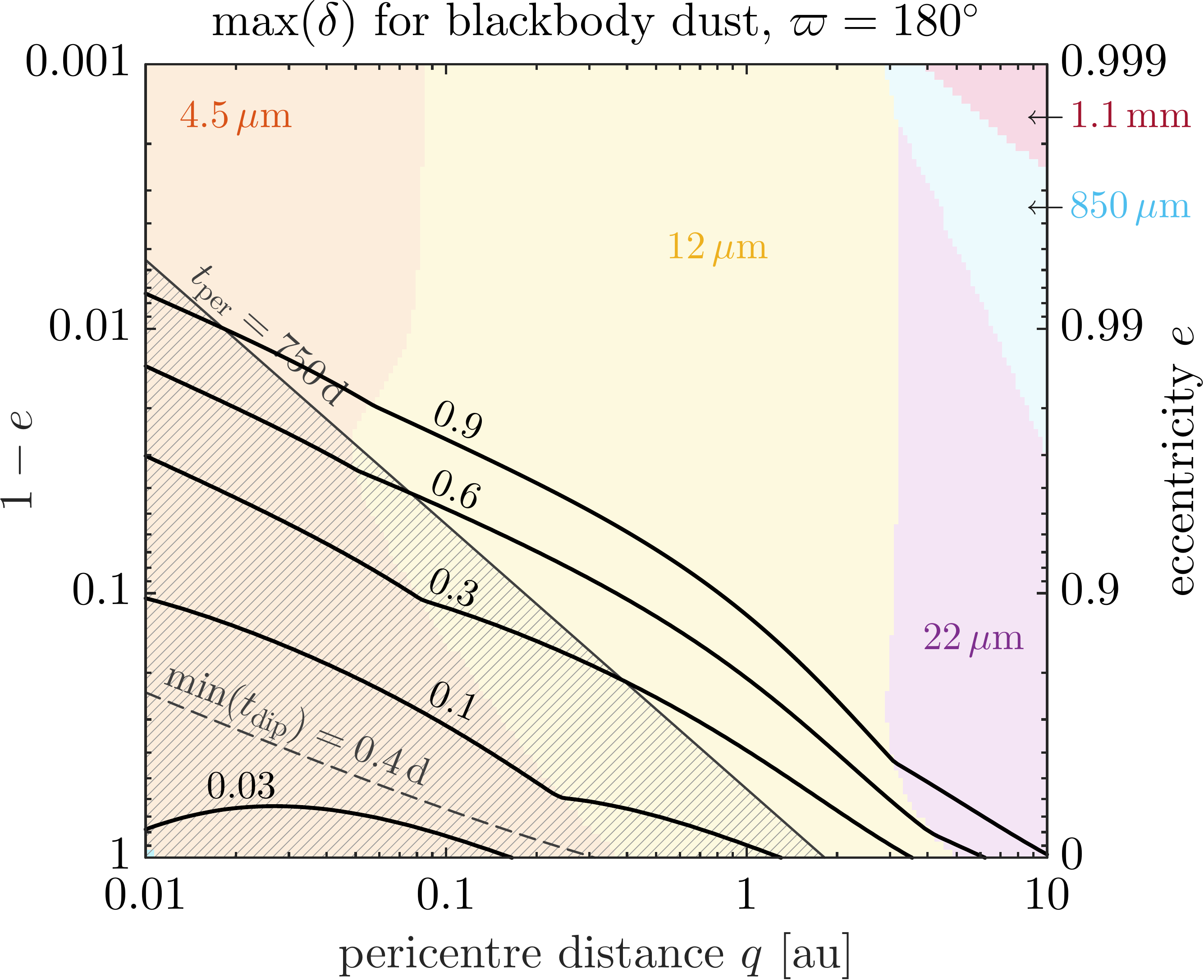} &
      \includegraphics[width=0.95\columnwidth]{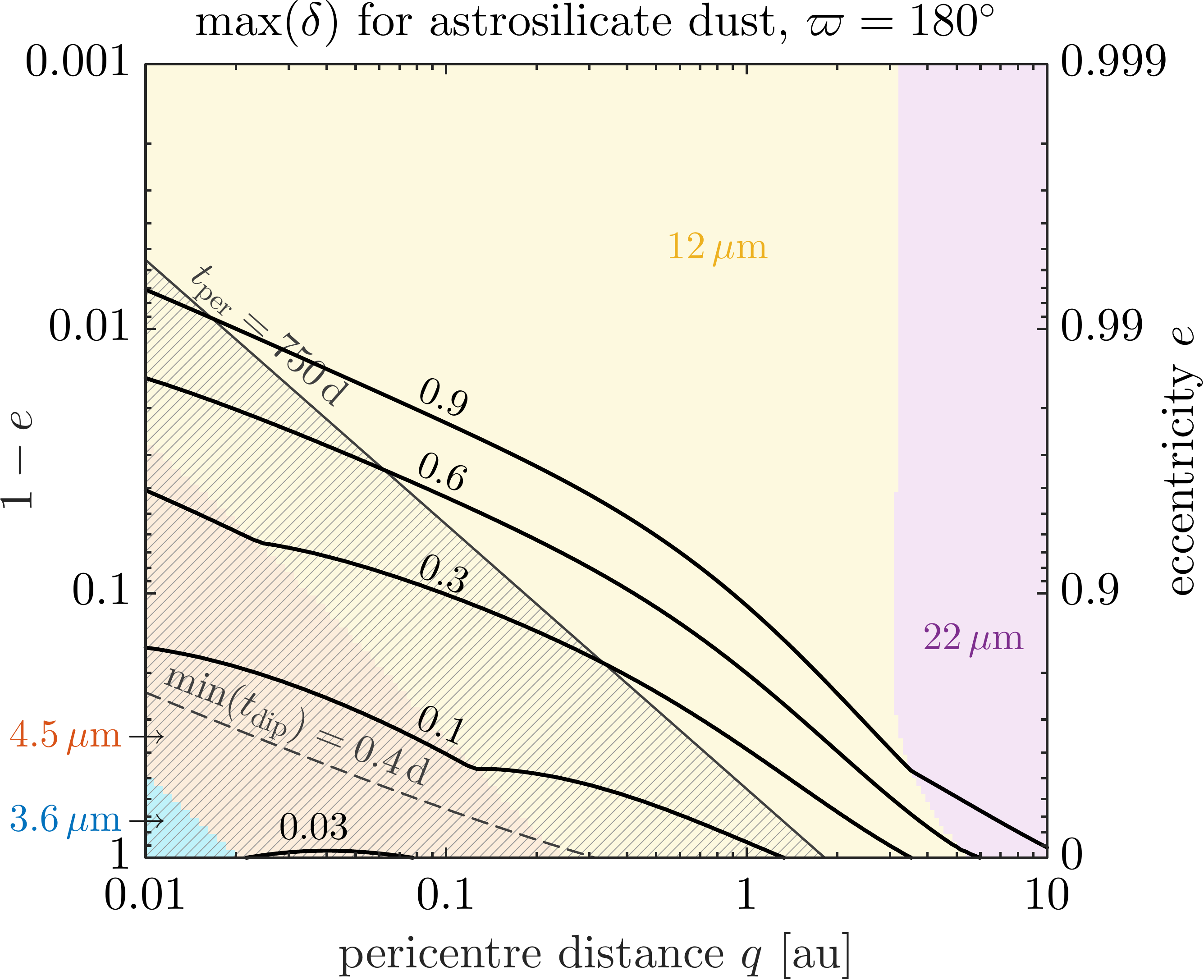}
    \end{tabular}
    \caption{Maximum possible level of dimming given the infrared observational constraints for
    different assumptions about the orbit, and about the grain optical properties.
    The top, middle and bottom panels show different assumptions about the orientation of
    the pericentre to the line of sight given by $\varpi$.
    The left panels assume black body grains, while the right panels assume grains with 
    the optical properties and size distribution described in \S \ref{ss:dust}.
    The solid lines show the maximum dimming allowed for the given orbital parameters, 
    which is at the level given in the annotation.
    The shading on the figure indicates the wavelength which provides the most stringent
    constraint on the level of dust for this orbit.
    The bottom left corner with the darker shading is that ruled out by the lack of
    repeating transits in a 750\,day timeframe.
    The orbit must lie to the left of the dashed line to allow dips as short as 0.4\,days.
    This plot assumes that the level of dimming does not change with time,
    that the stellar emission has been subtracted perfectly,
    and uses the approximation for optically thick distributions that the absorbed
    radiation is emitted isotropically (assumptions which are challenged in \S \ref{s:dist}, \ref{s:stellardim}
    and \ref{ss:thickedge}, respectively).
    }
   \label{fig:ir1}
  \end{center}
\end{figure*}

In Fig.~\ref{fig:ir1} we consider the emergent spectrum for different orbital configurations (pericentre
distance $q$, eccentricity $e$, and orientation of pericentre $\varpi$).
For the above assumptions the shape of this emergent spectrum is independent
of the optical depth of the distribution.
Thus, for each configuration, the fractional luminosity of the emission spectrum $f$ (i.e., the flux density
integrated over all wavelengths divided by the stellar luminosity) is increased until any of the upper
limits in Table~\ref{tab:fir} is reached.
The shading in Fig.~\ref{fig:ir1} shows the wavelength of observation which is most constraining for this orbit.
For optically thin distributions, the derived fractional luminosity would constrain
$\langle Q_{\rm abs} \rangle_{D,\star} f_{\rm BB}$ (see eq.~\ref{eq:f2}).
However, more generally equation~\ref{eq:fnew} shows that the derived fractional luminosity
constrains the combination
$\kappa[1-\exp{(-\langle\tau_{\rm abs}\rangle_\star)}]$ required to get this level of
dust emission.

The optical depth and annulus height (and width) are not uniquely constrained, but (when optical depth in absorption is
converted to that in extinction) these two parameters are all that is needed to give the level of
dimming via eq.~\ref{eq:deltadef} (and eq.~\ref{eq:omstar}).
In the limit of small optical depth (or large annulus height), $\Omega_\star$ is unity,
and so the resulting dimming scales linearly with optical depth (or inversely with annulus height).
In the limit of large optical depth (or small annulus height), $\Omega_\star \propto \kappa$
(see eq.~\ref{eq:omst}), and so the resulting dimming scales inversely with optical depth (and linearly with annulus height).
Clearly at some intermediate level of optical depth (or annulus height) the level of dimming for this configuration has a maximum
possible value, and that is what is assumed in Fig.~\ref{fig:ir1}.
For the above assumptions the optical depth is sufficiently small for it to be valid to calculate the
emission spectrum using equations~\ref{eq:fnu}-\ref{eq:tbb}.
However, we will return to optical depth effects in \S \ref{ss:thickedge}.

One conclusion from Fig.~\ref{fig:ir1} is that the preliminary conclusion from \S \ref{sss:dfrel}
that the material must be transiting beyond $\sim 1.5$\,au to cause secular dimming at a level
of 10\% (see eq.~\ref{eq:rmin}), is broadly correct.
This is most readily apparent from the fact that the lines of constant ${\rm max}(\delta)$
are mostly vertical in Fig.~\ref{fig:ir1} for the assumption that the transit occurs at pericentre.
However, this same conclusion also holds for transits at apocentre and in between.
For example, for transits at apocentre the lines of constant ${\rm max}(\delta)$ are sloping
in a way that keeps $r_{\rm t}$ approximately constant, since for this orientation and large
eccentricities $1-e \approx 2q/r_{\rm t}$.
It also holds (broadly) for both the black body and realistic grain assumptions
(i.e., left and right panels of Fig.~\ref{fig:ir1}, respectively).
The more detailed structure of the lines of constant ${\rm max}(\delta)$ on Fig.~\ref{fig:ir1},
however, improve on eq.~\ref{eq:rmin} (which only considered constraints on the total luminosity
of the dust) by taking account of the details of the emission spectrum and
the upper limits at different wavelengths.

The fact that the ${\rm max}(\delta)=0.1$ line lies in the yellow region shows that the most
constraining observation for this level of dimming is that at 12\,$\mu$m.
This will allow \S \ref{s:dist} to focus on when the observations at this wavelength were
made to consider whether the infrared constraints are compatible with the secular dimming.
The longer wavelengths are not so constraining, unless large levels of dimming ($>30$\%) are
required (which we know from eq.~\ref{eq:rmin} means the dust must be further from the star and so
colder).
Even then the millimetre wavelength observations do not provide significant constraints when realistic grain
properties are taken into account (see Fig.~\ref{fig:spec}).
The shorter (near-IR) wavelengths are not as constraining since they require dust to be very
close to the star, whereas the fractional luminosity constraints of eq.~\ref{eq:rmin} show that
the dust is passing the star at least 1.5\,au away at the point of transit.
The shape of the spectrum is also such that the 12\,$\mu$m observation is usually more
constraining unless all of the orbit is close to the star (see Fig.~\ref{fig:spec}).

Also plotted on Fig.~\ref{fig:ir1} is the constraint that the lack of
repeatability of any of the dips in the {\it Kepler} data implies an orbital period longer than
750\,days.
This is half of the period constraint used in \citet{Boyajian2016} to acknowledge the possibility
that the dip seen in 2017 May (Boyajian et al., in prep.) originates in the same material as that
seen in the D800 and D1500 dips of the {\it Kepler} data with a $\sim 750$\,day period
\citep[as proposed in section 4.4.3 of][]{Boyajian2016}.
The bottom panels of Fig.~\ref{fig:ir1} show that this is already inconsistent with witnessing the
transit at apocentre, if the same orbit has to explain secular dimming at a 10\% level.

\subsection{Short dip constraints for all dust distributions}
\label{ss:shortdip}
As noted in \citet{Boyajian2016} another constraint on the orbit comes from the duration of the short term dimming events
(i.e., the dips).
The most constraining of these is the shortest dip which lasted 0.4\,days (see \S \ref{s:lc}).
For a given vertical distribution of material (defined by $i$ and $\kappa$), this requires the material
to be moving at a minimum tranverse velocity $v_{\rm t}$, which for a given
orientation of the pericentre $\varpi$ and eccentricity $e$ translates into a maximum pericentre distance $q$.
This constraint is noted on Fig.~\ref{fig:ir1} for the assumption that $i=0$, i.e., that the dip duration is
the time it takes to cross $2R_\star$ and so $t_{\rm dip} < t_{\rm cross}(b=0) = 2R_\star/v_{\rm t}$.
Since a 0.4\,day dip must originate from an orbit to the left of this line, this is already inconsistent with an
orbit that also results in 10\% dimming given the infrared constraints of \S \ref{ss:irobs} regardless of the
pericentre orientation.

Another way of presenting this constraint is to write the crossing time in terms of the distance of the material
from the star at transit,
\begin{equation}
  t_{\rm cross}(b=0) = \left( \frac{2R_\star}{\sqrt{GM_\star}} \right)\sqrt{r_{\rm t}/(1+e\cos{\varpi})}.
  \label{eq:tcross}
\end{equation}
For KIC8462852, the existence of a 0.4\,day dip means that
\begin{equation}
  r_{\rm t} < 0.31 (1+e\cos{\varpi})
  \label{eq:rminsd}
\end{equation}
in au, and so $r_{\rm t} < 0.62$\,au.
This emphasises that the shortest duration dip is already in tension with the fractional
luminosity constraint of eq.~\ref{eq:rmin}, though the comparison is best done using
Fig.~\ref{fig:ir1}.
It also shows that transits oriented toward pericentre are favoured to reduce this tension.

\begin{figure}
  \begin{center}
    \begin{tabular}{c}
      \includegraphics[width=1.0\columnwidth]{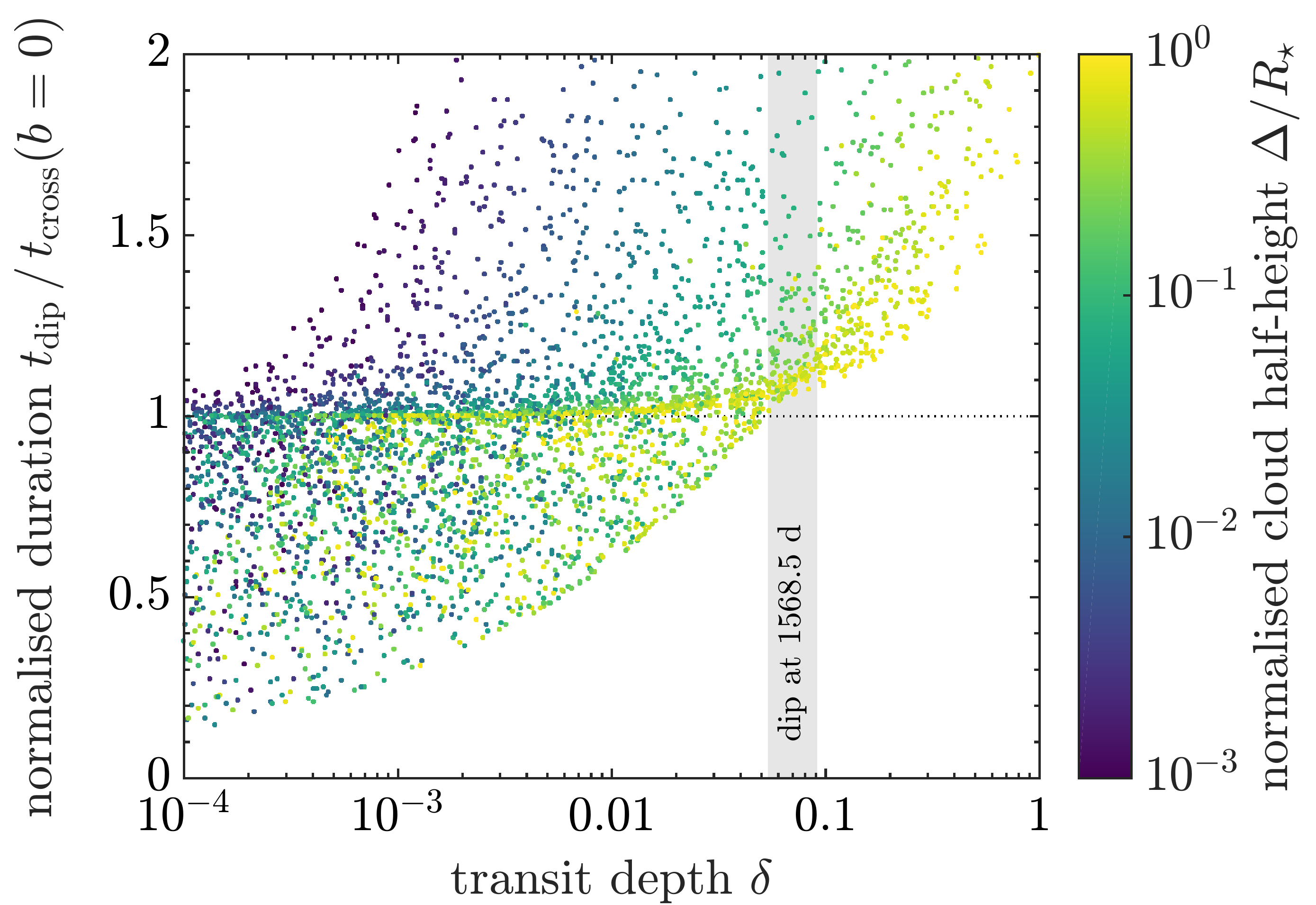}
    \end{tabular}
    \caption{Transit durations and depths for dust clouds with different
    heights and transit impact parameters.
    The transit duration is normalised to the time needed for a point-like
    particle to cross the $2R_\star$ diameter of the star.
    Each dot corresponds to a possible dust cloud described by the geometry
    introduced in \S~\ref{ss:geom}.
    The half-height of the cloud is picked from a log-uniform
    distribution between $10^{-3}R_\star$ and $1R_\star$, as indicated
    by the colour of the dot.
    Impact parameters are distributed uniformly between 0 and $2R_\star$.
    The rectangular dust cloud is assumed to be optically thick.
    A shaded vertical band indicates the approximate depth of the shortest
    dip found in the \textit{Kepler} light curve (see Fig.~\ref{fig:lczoom}).}
   \label{fig:sd}
  \end{center}
\end{figure}

We could try to argue that the material that causes the short duration dip and that causing the dimming are on
different orbits, but this is disfavoured by the improbability of witnessing events from two different orbits in
the same system, unless the some mechanism favours placing debris on two orbits that are aligned with our line of sight.
It might also be hoped that the dip duration constraint is conservative because viewing orientations with $i>0$
would result in shorter dips.
However, Fig.~\ref{fig:sd} shows that the fact that the observed 0.4\,day dip has a relative depth of $\delta=0.07$
precludes this helping this constraint much, since such a large depth requires a certain fraction of the stellar
disk to be blocked.
At shallow transit depths, transit durations can indeed be short;
e.g., for $\delta<0.06$ the lower bound on the transit duration as function of depth in Fig.~\ref{fig:sd} is set by grazing
dust clouds that block a small patch of the stellar disk next to its limb and so can cross quickly.
For higher depths $\delta>0.06$, however, the lower bound is set by vertically elongated clouds that block a strip of the
stellar disk from pole to pole, for which the dips are if anything slightly longer than given by eq.~\ref{eq:tcross}.
For transit depths below $\delta \lesssim 0.1$, there is a concentration of normalised
transit durations around unity caused by the shape of the stellar disk, since
for small dust clouds a relatively wide range of different impact parameters yields
normalised transit durations close to unity, which is thus a reasonable approximation
for the analysis.

There are four possible solutions to the problem of short dips (when trying to explain these
with the same circumstellar material causing the secular dimming):

(1) Ignoring the century duration secular dimming, the infrared constraints would only have to allow dimming up
to 3\% as seen by {\it Kepler} \citep{Montet2016}, which for the assumptions thus far is compatible with the
0.4\,day dip duration for an orbit with $q=0.4-0.6$\,au observed at
pericentre.
An orbit with $q=0.1-0.2$\,au may be allowed if the orientation is $\varpi=90^\circ$ (depending on the detailed
particle composition), but a viewing orientation along apocentre is not allowed. 

(2) The dust could be unevenly distributed around the orbit and so the infrared excess would be expected to vary
with time, which would have to be taken into account in the interpretation of the infrared observations.
This possibility is explored in \S \ref{s:dist}.

(3) Thus far we have only considered the constraints on the thermal emission, which assumes that it has been
possible to subtract the contribution of stellar emission from the observed flux density.
While it is possible to measure the optical brightness of the star at the epoch of the infrared measurements
to aid this subtraction (see, e.g., Meng et al., submitted), the stellar emission will be subject to a
different level of extinction at infrared wavelengths (see \S \ref{ss:dust}).
Oversubtraction of the stellar emission could make it appear that there is less thermal emission than
is the case resulting in overly stringent constraints.
This is discussed in \S \ref{s:stellardim}.

(4) While the analysis has accounted for many optical depth effects, one limitation is that the energy
that is absorbed is assumed to be reradiated isotropically (and at the temperature expected for optically thin dust).
For optically thick distributions, the emission is generally lower for edge-on orientations than for face-on
orientations.
This could mean that the fractional luminosity observed is lower than that of the
whole disk.
That is, the ${\rm max}(\delta)$ constraints in Fig.~\ref{fig:ir1} could be underestimated because larger
levels of observed dimming are possible with the given configuration if the emission reaching us has been
attenuated by optical depth effects.
This effect is considered further in \S \ref{ss:thickedge}.

\section{Uneven dust distribution around the orbit}
\label{s:dist}
For an assumed orbit, we can use a similar method to that in Fig.~11 of \citet{Boyajian2016}
to convert the observed light curve into the distribution of dust around the orbit required
to reproduce that light curve, which can then be converted
into a prediction for the evolution of infrared excess versus time.
The method is described in \S \ref{ss:method2} and applied in \S \ref{ss:irvt} to the
KIC8462852 light curves discussed in \S \ref{s:lc}.

\subsection{Converting optical dimming to infrared excess}
\label{ss:method2}
For a given orbit ($q$, $e$) and pericentre orientation ($\varpi$),
we know the velocity at which the material crosses the star $v_{\rm t}$ (eq.~\ref{eq:vt}).
For a given orbital inclination ($i$) and distribution about the reference orbit ($\kappa$),
it is possible to determine how an element of material of projected length along the orbit $ds$
(e.g., the dark shaded element in Fig.~\ref{fig:geom} right), with a cross
sectional area $d\sigma$, affects the lightcurve, since material at impact parameters
$b$ to $b+db$ causes an upside down top-hat shaped dip with duration $2v_{\rm t}^{-1}\sqrt{R_\star^2-b^2}$
and depth that depends on the optical depth of the element $\tau_{\rm ext}$.

Here we are assuming that elements have uniform optical depth as a function of impact parameter (from
$b_{\rm l}$ to $b_{\rm u}$, see Fig.~\ref{fig:geom} right).
Thus the optical depth of the element that passes through transit at time $t$ is given by
\begin{equation}
  \tau_{\rm ext}(t) = \langle Q_{\rm ext} \rangle \dot{\sigma}(t) / (2 \kappa h),
  \label{eq:tauextt}
\end{equation}
where $\dot{\sigma}(t)$ is the rate at which cross-sectional area passes transit at time $t$.
Integrating over all transit times gives for the level of dimming as a function of time
\begin{equation}
  \delta(t) = \int [1 - \exp{(-\tau_{\rm ext}(t'))}] H(t-t') 2 \kappa h dt' / (\pi R_\star^2),
  \label{eq:deltat}
\end{equation}
where $H(t-t')$ is the fraction of the element that is in front of the star at a time $t-t'$ after
transit.

While equation~\ref{eq:deltat} can be used to infer the optical depth due to extinction as
a function of time (and so through equation~\ref{eq:tauextt} the rate at which cross-sectional
area crosses the star as a function of time), this inference is complicated by the fact that
the cross-sectional area evolution has been convolved with the $H(t-t')$ function. 
That is, there is no unique solution at the resolution of the time it takes for an element
to cross the star.
Consideration of that convolution will be important for interpreting the shape of 
the short duration dips.
However, for the interpretation of the secular dimming, and of all but the shortest
duration dips, the cross-sectional area is not changing rapidly, and so
equation~\ref{eq:deltat} can be rewritten as
\begin{equation}
  \delta(t) \approx [1 - \exp{(-\tau_{\rm ext}(t))}]\Omega_\star,
  \label{eq:deltat2}
\end{equation}
where we have used $\Omega_\star = \int H(s) 2 \kappa r_{\rm t} ds / (\pi R_\star^2)$.
This can be inverted and combined with equation~\ref{eq:tauextt} so that the light curve can be used to find
\begin{equation}
  \dot{\sigma}(t) \approx -\left( \frac{2 \kappa h}{\langle Q_{\rm ext} \rangle} \right) \ln{[1-\delta(t)/\Omega_\star]}.
\end{equation}
For the case that the distribution is narrow so that $\kappa \ll R_\star/r_{\rm t}$, and optically thin so
that $\delta(t)/\Omega_\star \ll 1$,
this gives for an edge-on orbit ($i=0$) that $\dot{\sigma}(t) \approx \delta(t) (\pi/2) h (R_\star/r_{\rm t}) / \langle Q_{\rm ext} \rangle$.

Having derived $\dot{\sigma}(t)$, albeit given assumptions about the orbit, it is then possible to determine
the expected thermal emission as a function of time.
For each element $d\sigma=\dot{\sigma}(t')dt'$ that passes at time $t'$ in a timestep $dt'$, its distance
from the star as a function of time $r(t-t')$ can be determined from the fact that the true anomaly
(i.e., the angular distance from pericentre along the orbit) $\theta=-\varpi$ at the point of transit (see Fig.~\ref{fig:geom} left).
The linear increase in mean anomaly with time and Kepler's equation can then be used to get $\theta(t-t')$
and so $r(t-t')$ \citep[see, e.g.,][]{Murray1999}.
The thermal emission from the element as a function of time at different wavelengths then comes from similar
reasoning to that which arrived at equations~\ref{eq:fnu}-\ref{eq:tbb}.
Summing over all elements we find that
\begin{equation}
  F_{\nu,{\rm th}}(t) = \int G(t-t') \dot{\sigma}(t') dt',
  \label{eq:fnut}
\end{equation}
where
\begin{eqnarray}
  G(t-t') & = & 2.35 \times 10^{-11} d^{-2} \int_{D_{\rm min}}^{D_{\rm max}} Q_{\rm abs}(\lambda,D) \times \nonumber \\
          &   & B_\nu(\lambda,T[D,r(t-t')])\bar{\sigma}(D)dD    
  \label{eq:gtt}
\end{eqnarray}
is the thermal emission per cross-sectional area as a function of time $t$ for the element
that passed through transit at time $t'$, and $\bar{\sigma}(D)dD$ is the fraction of its cross-sectional area
in sizes from $D$ to $D+dD$.
 
Equations~\ref{eq:fnut} and \ref{eq:gtt} implicitly assume that the emission is optically thin, and further
corrections are needed for optically thick emission.
For the assumptions made so far, that the optically thick emission is isotropic and has the
same temperature as optically thin emission, this could be accounted for by an additional factor of
$ [ 1 - \exp{ ( -\langle \tau_{\rm abs}\rangle_\star(t') ) } ] / \langle \tau_{\rm abs}\rangle_\star(t') $
inside the integral in equation~\ref{eq:fnut}.
However, there would be additional complications due to the geometrical effects discussed in \S \ref{ss:thickedge}
where an alternative correction is proposed.

\subsection{Infrared flux versus time for KIC8462852}
\label{ss:irvt}

\begin{figure*}
  \begin{center}
    \begin{tabular}{c}
      \includegraphics[width=2.0\columnwidth]{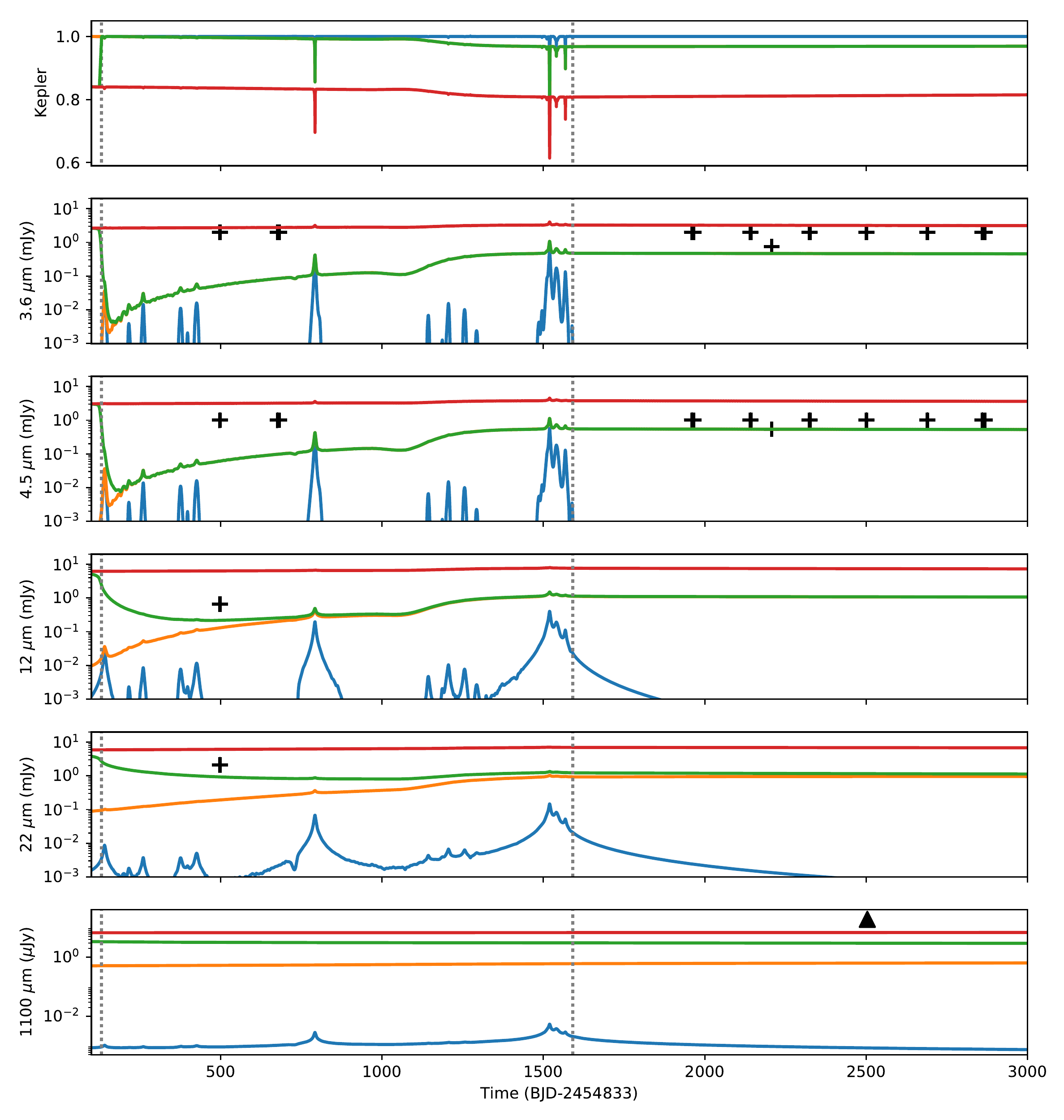}
    \end{tabular}
    \caption{Based on the different interpretations of secular dimming for the optical light curves
    from \S \ref{s:lc} that are reproduced in the top panel, the other panels show the predicted light curves
    for the infrared thermal emission (i.e., ignoring the contribution of stellar emission at these wavelengths).
    The $x$-axis is time from 2009 April until 2017 March; the vertical dashed lines demark the range of times
    when {\it Kepler} observed the star.
    An orbit with $q=0.1$\,au, $e=0.9975$ ($a=40$\,au), and $\varpi=0^\circ$ (i.e., viewed along pericentre)
    is assumed.
    The dust optical properties and size distribution from \S \ref{ss:dust} are used, and optical depth
    effects included for the assumption that the emission from optically thick distributions is isotropic
    and at the same temperature as for optically thin dust
    (see \S \ref{ss:thickedge} for a more realistic consideration of optically thick distributions). 
    The different colours correspond to the different assumptions about the secular dimming
    (i.e., blue is no dimming, orange is {\it Kepler}'s 3\% dimming, red includes 16\% century-long dimming,
    green includes 16\% century-long dimming which disappeared before {\it Kepler} started observing).
    The pluses show upper limits from {\it WISE} and {\it Spitzer} from Table~\ref{tab:fir}.
    The upper limit at 1.1\,mm is far off the top of the plot at the epoch indicated by the triangle.
    }
   \label{fig:irvt}
  \end{center}
\end{figure*}

Here we apply the method of \S \ref{ss:method2} to the light curves of KIC8462852 with different assumptions about
the secular dimming (see \S \ref{s:lc}) to make predictions for the infrared flux as a function of time.
The optical light curves are reproduced in the top panel of figure~\ref{fig:irvt}, while the panels below
give the predicted infrared flux curves, along with the infrared upper limits from Table~\ref{tab:fir}
\citep[noting that while 3.4 and 4.6\,$\mu$m observations continue to be taken with the {\it NEOWISE} Reactivation,]
[12 and 22$\mu$m measurements were only taken once, during the initial cryogenic phase]{Mainzer2011}.
As noted in \S \ref{ss:method2}, this prediction requires an assumption about the orbit, which for reasons that
will become clear in \S \ref{sss:revised} has a pericentre $q=0.1$\,au and eccentricity $e=0.9975$ (and so a semimajor axis $a=40$\,au
and an orbital period of 211\,yr).
This orbit is assumed to be viewed edge-on ($i=0$), with its pericentre along the line of sight
($\varpi=0^\circ$).
The size distribution and realistic optical properties of \S \ref{ss:dust} are also assumed, and the
distribution is assumed to be optically thin.

\subsubsection{Revised constraints from secular dimming}
\label{sss:revised}
When discussing the predictions of Fig.~\ref{fig:irvt} it should first be acknowledged that the analysis in
\S \ref{ss:irobs} already provides an accurate estimate of the expected flux level for situations
in which the level of dimming is constant (or not changing significantly).
For example, the top right panel of Fig.~\ref{fig:ir1} shows that for orbits with $q=0.1$\,au and $1-e=0.0025$,
the maximum level of dimming that is allowed before breaking the 12\,$\mu$m upper limit is $\sim 1$\%.
This explains why Fig.~\ref{fig:irvt} finds a predicted 12\,$\mu$m flux level for the red light curve that is an order
of magnitude larger than the upper limit, since this involves a level of dimming that
is constant at 16-19\% for a long period.
The first conclusion therefore, is that the 12\,$\mu$m upper limit rules out that the century-long dimming
persisted until the time of {\it Kepler} (although we will revisit this in \S \ref{ss:thickedge}).

\begin{figure*}
  \begin{center}
    \begin{tabular}{c}
      \includegraphics[width=2.0\columnwidth]{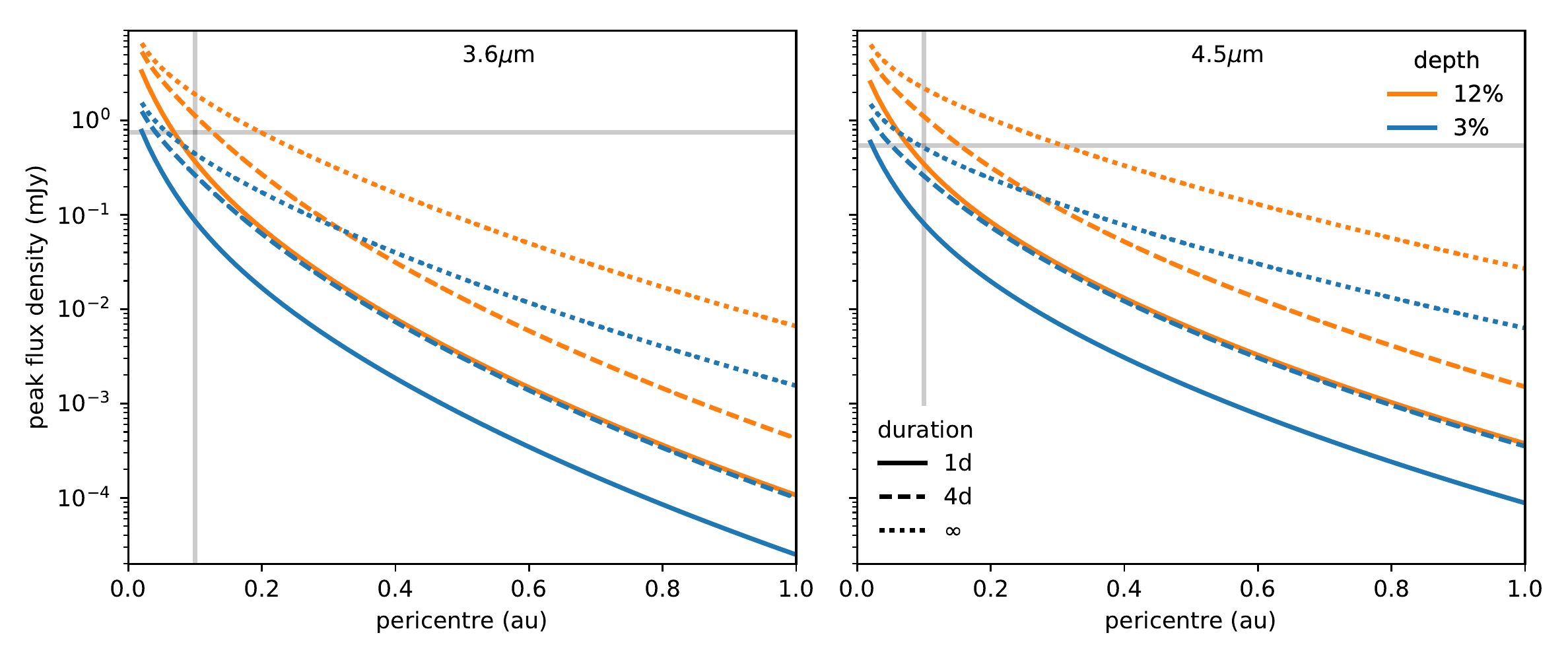}
    \end{tabular}
    \caption{Peak infrared thermal emission (at 3.6\,$\mu$m left and at 4.5\,$\mu$m right)
    as a function of pericentre distance for a dimming event seen in the
    optical light curve with a depth of 3 and 12\% shown in blue and orange, respectively.
    The orbit is assumed to be parabolic and aligned with its pericentre along the line of sight ($\varpi=0$).
    The dimming event is assumed to be square-shaped and of durations 1\,day, 4\,days or $\infty$ (i.e., constant
    dimming level) shown with solid, dashed and dotted lines, respectively.
    The dust optical properties and size distribution from \S \ref{ss:dust} are used, and optical depth
    effects included for the assumption that the emission from optically thick distributions is isotropic
    and at the same temperature as for optically thin dust
    (see \S \ref{ss:thickedge} for a more realistic consideration of optically thick distributions). 
    The {\it Spitzer} upper limits from Table~\ref{tab:fir} at these wavelengths are shown with horizontal
    solid grey lines, and the vertical grey lines show the pericentre distance of $q=0.1$\,au from Fig.~\ref{fig:irvt}.
    }
   \label{fig:irvq}
  \end{center}
\end{figure*}

Applying the same comparison to the other light curves (i.e., the other interpretations of the secular dimming), however,
finds that the 12\,$\mu$m upper limit is not as constraining as had been concluded in \S \ref{ss:irobs}.
This is because, for the assumed orbit, the 12\,$\mu$m emission is dominated by warm material that is not far from pericentre
and so must have passed across our line of sight within $\sim 100$\,days of the measurement.
Since the secular dimming is much smaller than 1\% at the epoch the 12\,$\mu$m measurement was made for all of
the other light curves, this measurement does not exclude the assumed orbit as might have been inferred from Fig.~\ref{fig:ir1}.
The second conclusion therefore, is that as long as the century-long dimming had disappeared by the
time {\it Kepler} started observing, the strongest constraints on the secular dimming come from the near-IR measurements that were
made at an epoch when we can be confident that there was secular dimming at the $\sim 3$\% level.
Indeed, this was how the orbital parameters were chosen, so that the predicted flux at $\sim 4.5$\,$\mu$m for
{\it Kepler}-only secular dimming is coincident with the most stringent upper limit at that wavelength at that epoch
(which is that from {\it Spitzer}).

While the above discussion focussed on the predictions for an orbit with $q=0.1$\,au and $e=0.9975$, the conclusions that
were reached about what can be learned from the 12\,$\mu$m upper limit apply more generally.
This is because we have already concluded from the shortest dips that the material must be transiting within
1\,au from the star, which means that regardless of the orbit, it is only the material that causes dimming at
epochs within a few 100\,days that contributes to the 12\,$\mu$m emission at a particular epoch.
In the absence of the 12\,$\mu$m constraint, the lines of maximum dimming would move to the left on
Fig.~\ref{fig:ir1} and the region over which the 4.5\,$\mu$m upper limit becomes the most constraining would
expand to encompass orbits previously constrained by the 12\,$\mu$m observation. 

To quantify this new constraint on the pericentre distance, Fig.~\ref{fig:irvq} shows the near-IR 
flux that would be expected for material that causes dimming at a given level when observed toward
pericentre.
This shows that the pericentre must be $q>0.1$\,au for the 3\% secular dimming seen by {\it Kepler} to be consistent with
the 4.5\,$\mu$m upper limit (i.e., where the dotted blue line on the right hand plot is below the horizontal grey line). 
For different pericentre orientations, the flux level would still be determined by the pericentre
distance, but the level of dimming would be larger than shown on Fig.~\ref{fig:irvq} due to the 
vertically broader dust distribution for a point of transit that is further from pericentre. 
This means that for such orientations, smaller pericentres can accommodate both 3\% secular dimming
and the 4.5\,$\mu$m constraint;
for example, for $\varpi=90^\circ$ the dimming is consistent with the flux constraint for pericentres
beyond 0.04\,au which means that the transit occurs at $r_{\rm t}>0.08$\,au (i.e., there is
little change to the constraint on the distance of material from the star at the point of transit
due to a change in pericentre orientation).
Thus we conclude that the combined constraints from the infrared, dimming and short dip duration,
are that the material transits the star at a distance 0.1-0.6\,au, which is not necessarily through
pericentre. 

As already discussed in \S \ref{ss:irobs}, the use of realistic dust optical properties means
that the millimetre wavelength observations do not provide significant constraints on the proposed scenario.
However, it is worth noting from the bottom panel how the longest wavelengths are uniquely sensitive
to the century-long dimming.
For example, the predictions for the green and orange lines are mostly indistinguishable for near-IR and
mid-IR wavelengths, even though the green line includes a substantial level of dimming in the pre-{\it Kepler} era
which is absent for the orange line.
This is because the material causing the century-long dimming is already at large enough distances by the
{\it Kepler} epoch (which is that plotted) to emit very little at such short wavelengths.
In contrast, the long wavelength flux remains sensitive to this distant material, and so the predicted
flux level depends on the precise prescription for secular dimming outside the {\it Kepler} era.
While the predicted level in Fig.~\ref{fig:irvt} would be challenging even with {\it ALMA}, this could be higher
with a different dust model (e.g., with larger grains which would have a more black body-like spectrum,
see Fig.~\ref{fig:spec}).

\subsubsection{Short-lived increases in infrared flux}
\label{sss:spike}
Another aspect of the infrared light curve which is evident in Fig.~\ref{fig:irvt} is that there are
short-lived increases in brightness that are coincident with the dips in the optical light curve.
These short-term dimming events (i.e., dips) are inferred to arise from dust concentrations that are more discretised
than the broader dust level that is causing the secular dimming. 
The peaks in near- and mid-infrared brightnesses always occur when the material passes through pericentre.
Thus the optical dips and infrared peaks only coincide exactly when the orientation is such that
it is the pericentre that is seen at transit.
For different orientations $\varpi \ne 0$ the infrared peak would be offset from the optical dip by
an amount that depends on $\varpi$ (that can be worked out using Kepler's equation).
However, this offset is generally small, because for the high eccentricities considered here it is only along
viewing angles very close to the apocentre direction that particles move slowly in true anomaly, and so the
time taken for a particle to move from a point between us and the star (transit)
and pericentre is short.

The infrared peaks are broader than the optical dips because the material contributes to the infrared flux
at all points around the orbit, not just at transit.
For example, an element of material that causes a narrow dip is spread out in time by the convolution
given by $G(t-t')$ (see eq.~\ref{eq:gtt}).
The shape of this convolution can be inferred from Fig.~\ref{fig:irvt} by looking at the predicted
flux from the blue light curve near the short-lived dip at $\sim 800$\,days.
This shows that the dips are spread out in time more at longer wavelengths, which is because longer wavelengths
are sensitive to material out to larger distances from the star.

It is notable that for the assumed orbit the peaks in near-IR flux are close to the flux levels of the
observed upper limits.
Thus, while there were no dips (and so no predicted infrared peaks) at the epochs of the infrared observations,
this shows that the predicted flux level of the peaks is within current observational capabilities. 
The detection of such features would provide significant constraints on the orbit and dust properties.
In addition to the timing of the peak relative to the optical dip constraining the pericentre orientation,
the flux level is set by the pericentre distance in a way that is shown in Fig.~\ref{fig:irvq}.
For closer pericentre distances the peaks become more pronounced.
However, note that for short dips it is the product of the depth and duration which is constrained
(since this determines the total amount of material present in the dip).
This is clearest to see by noting that the predicted peak flux level is the same for the
3\% dip that lasts 4\,days as for the 12\% dip that lasts 1\,day.

\section{Total infrared flux: including stellar emission}
\label{s:stellardim}
Thus far the model has only been used to derive the thermal emission from the circumstellar dust at infrared
wavelengths and to compare that with the optical brightness of the star which has been subject to extinction by the same
material.
However, the total observed infrared flux includes a contribution both from the dust thermal emission and from
the stellar emission which has been subjected to a different level of extinction to that in the optical.
In general the total observed flux is
\begin{equation}
  F_{\nu,{\rm obs}}(\lambda) = F_{\nu,{\rm th}}(\lambda) + F_{\nu,\star}(\lambda) (1-\delta_\lambda),
  \label{eq:fnuobs}
\end{equation}
where the $(1-\delta_\lambda)$ factor can be derived from Fig.~\ref{fig:dldv} for a given level of optical
dimming $\delta_{\rm V}$ and assumptions about how optically thick the distribution is
(i.e., how vertically narrow it is).

The fluxes in Table~\ref{tab:fir} used the observed infrared fluxes to set constraints on the thermal emission.
For this calculation grey extinction was assumed ($\delta_\lambda=\delta_{\rm V}$), which is a reasonable 
assumption for large particles or optically thick distributions (see Fig~\ref{fig:dldv}).
However, even then it is not possible to get independent constraints on the thermal emission in this way, because
the thermal emission and optical brightness of the star are to some extent (anti-)correlated.
We illustrate this in \S \ref{ss:newdip} by considering the brightness evolution at different wavelengths
during a short-term dimming event, and then return to the implications for secular dimming in \S \ref{ss:newdim}.

\subsection{Application to 2\% dip}
\label{ss:newdip}

\begin{figure*}
  \begin{center}
    \begin{tabular}{c}
      \hspace{-0.5in} \includegraphics[width=1.6\columnwidth]{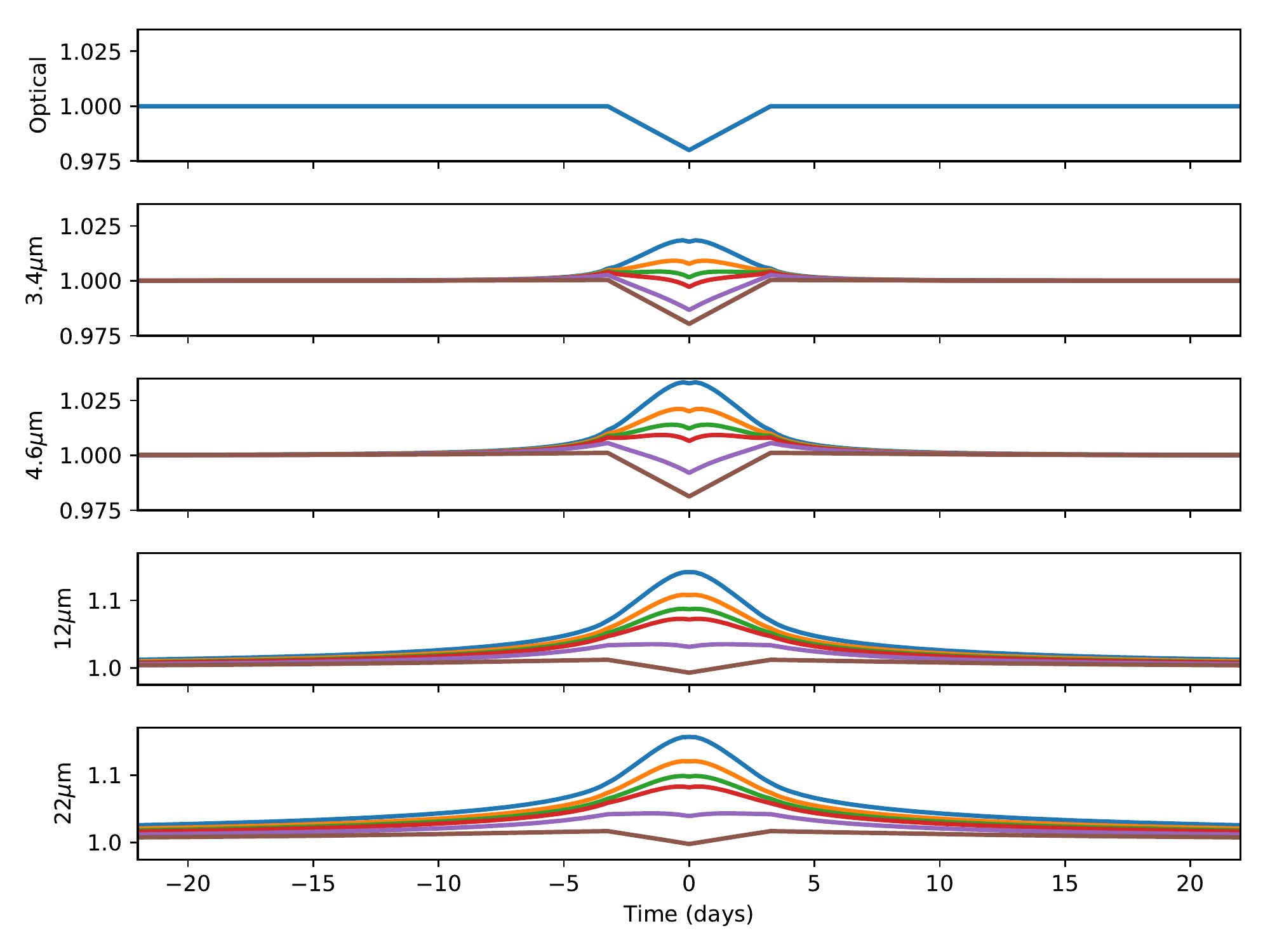} \\
      \includegraphics[width=1.7\columnwidth]{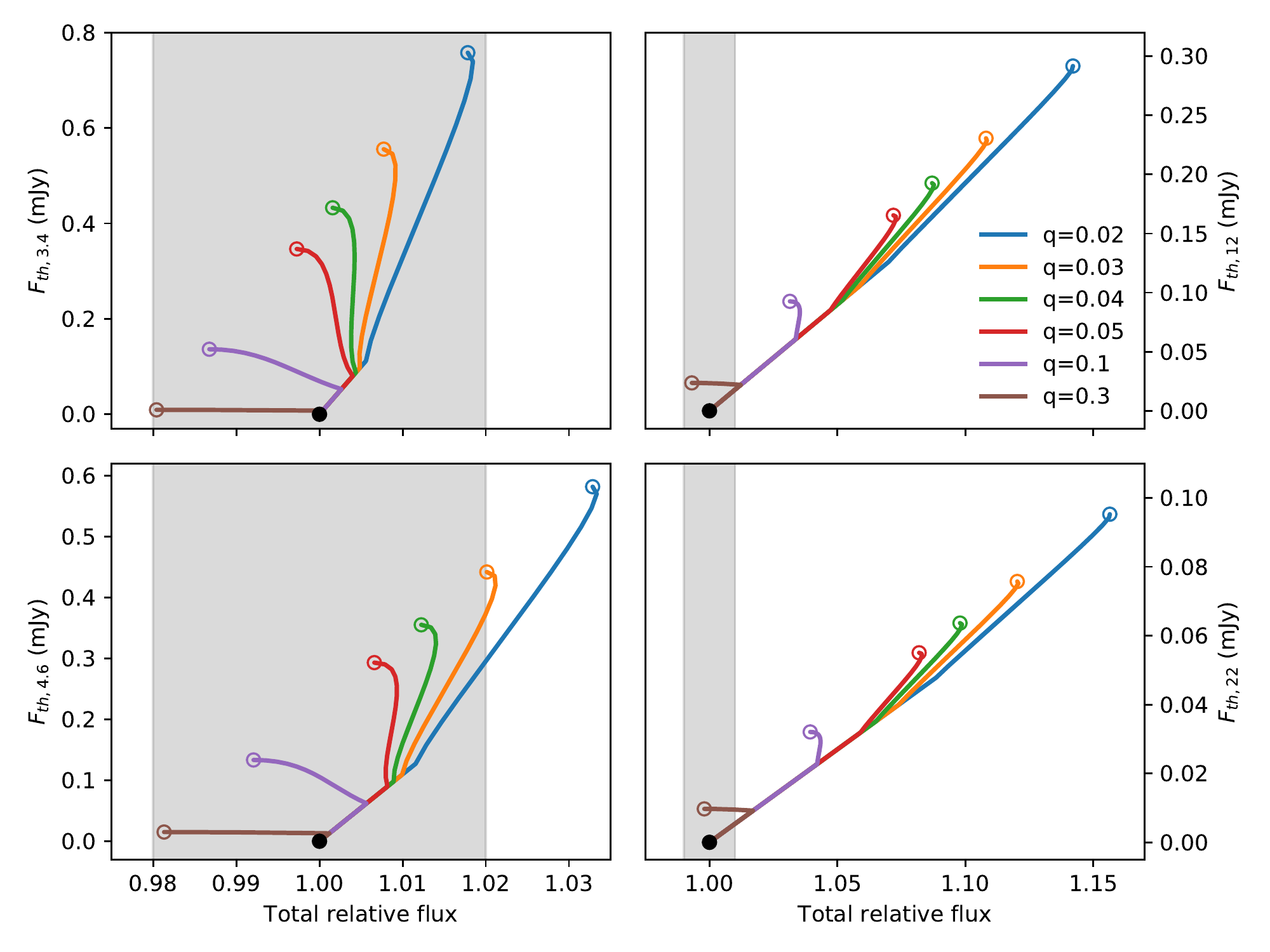}
    \end{tabular}
    \caption{Predicted total infrared fluxes (i.e., including both the dust thermal emission and the contribution
    of starlight) for a triangular-shaped 2\% dip in brightness
    of duration 6.5\,days, for an orbit observed towards pericentre $\varpi=0$.
    The dust optical properties and size distribution from \S \ref{ss:dust} are used, and optical depth
    effects included for the assumption that the emission from optically thick distributions is isotropic
    and at the same temperature as for optically thin dust
    (see \S \ref{ss:thickedge} for a more realistic consideration of optically thick distributions). 
    {\bf (Top)} Total observed flux as a function of time in near-IR and mid-IR, for
    pericentre distances of 0.02\,au (blue) to 0.3\,au (brown). 
    {\bf (Bottom)} Tracks that the light curves take at the different wavelengths,
    with the different pericentre distances shown in different colours, on a plot
    of the thermal emission from the material causing the dip against the total 
    observed flux.
    The observational constraints at the different wavelengths are shown by the grey shaded region,
    where these constraints come from {\it NEOWISE} at 3.4 and 4.6\,$\mu$m and are the constraints that
    may be provided by a future space-based mid-IR telescope at 12 and 22\,$\mu$m.
    The tracks start and end at the black filled circle when the material causing the dip is at apocentre,
    although the track only moves far from this point when the material is close to pericentre
    which is shown with the open coloured circle.
    }
   \label{fig:newdip}
  \end{center}
\end{figure*}

Continued monitoring of KIC8462852 has resulted in the discovery of a new dimming event which occurred
between 2017 May 18-24 with further dimming ongoing at the time of submission of this paper (Boyajian et al., in prep.).
The peak dip depth of the first event was $\sim 2$\%, the event lasted a few days, and
the integrated depth was $\sim 6.5$\%\,days (i.e., equivalent to a 1\% dip that lasted 6.5\,days).
Fortunately the dip (known among the discovery team as {\it Elsie}; Boyajian et al., in prep.)
occurred at an epoch at which {\it NEOWISE} \citep{Mainzer2011} was observing the star so the near-IR flux
of the star was also measured both immediately before and during the dipping event.
Since no increase in emission was detected by {\it NEOWISE} during the event, the resulting upper limit on the flux
is approximately the same as that of the previous {\it WISE} measurements in Table~\ref{tab:fir} (Boyajian et al., in prep.).
We estimate the 1$\sigma$ uncertainty at 3.4 and 4.6$\mu$m as 2\%, approximately the precision of a single-frame
{\it NEOWISE} Reactivation measurement.

In Fig.~\ref{fig:newdip} we consider the predictions of the model for the evolution of
total infrared brightness for the {\it Elsie} dip using the method of \S \ref{ss:method2} along with eq.~\ref{eq:fnuobs}.
For simplicity the dip is assumed to be triangular in shape with the same duration and integrated depth
as that observed (with no secular dimming).
The orbit is assumed to be observed along pericentre, with different pericentre distances shown by the different
coloured lines.
Fig.~\ref{fig:irvq} already showed how the peak flux density in thermal emission would be expected to increase as the
pericentre is decreased.
While this is also the case for the total flux for small pericentre distances or long wavelengths,
Fig.~\ref{fig:newdip} shows that the decrease in stellar emission during the dip can dominate the overall
change in observed emission for large pericentre distances or short wavelengths.

Generally, the shape of the light curve at all wavelengths is an increase in flux that is centred on
the optical dip (because of the assumption that $\varpi=0$), which starts before and ends after the
dipping event (because material also contributes to the thermal emission when it is not in front of the star),
with a triangular shaped dip coincident with the optical dip.
The evolutionary track of the light curves is also shown in the bottom panels of Fig.~\ref{fig:newdip},
in which the thermal emission starts very close to zero and the total relative flux at unity when the
material is close to apocentre (i.e., at the black circles on these panels).
As the material approaches transit the thermal emission increases, as does the total flux proportionally,
until the material starts to pass in front of the star.
Thereafter the infrared flux continues to rise while the total flux either increases at a lower level, or can also
reverse that trend and decrease, until mid-transit (which is when the material is at pericentre and so located
at the open circles on Fig.~\ref{fig:newdip} bottom). 
Because we assume an event with a symmetric light curve, the track after mid-transit is the reverse of that
pre-transit, until the material again reaches apocentre.

For the near-IR panels of Fig.~\ref{fig:newdip} bottom, the grey bars show the approximate constraints on the
total flux from the 2017 {\it NEOWISE} observations.
These show that, for similar reasons to the conclusion in \S \ref{sss:spike},
the peak 4.6\,$\mu$m flux rules out orbits with pericentres that are inside 0.03\,au (see the orange
line on this figure).
The mimimum 3.4\,$\mu$m flux also appears to rule out orbits with pericentres that are
outside 0.3\,au (see the brown line on this figure).
However, for pericentres beyond 0.3\,au, the thermal emission at 3.4\,$\mu$m is negligible
and so the decrease in relative flux is at most that of the optical dip depth plus a
correction for the wavelength dependence of the extinction (i.e., the track for pericentres beyond
0.3\,au look very similar to that for 0.3\,au).
Thus for a dip of this magnitude no constraints can be set from the absence of dimming seen at 3.4\,$\mu$m.
Such a constraint would have been possible had the dip been of greater depth
(or the observations of greater sensitivity), although 
it would still be important to be cautious with such an interpretation, because the
extinction in the model is assumed to be essentially grey (see Fig.~\ref{fig:dldv}), which may
not be the case. 
Had the dust been assumed to be made up of smaller particles then this would have had little
effect on the thermal emission but would have resulted in a decreased depth of infrared dimming
relative to that seen in the optical.

Deep ($\lesssim$1mJy) and high precision ($\sim$1\%) mid-IR measurements are not possible from the ground.
Thus, with no space telescope currently operational in the mid-IR, the constraints on the 12-22\,$\mu$m
observed fluxes during the dip are not close to the levels shown on Fig.~\ref{fig:newdip}.
Instead the grey bars on the 12 and 22\,$\mu$m panels of Fig.~\ref{fig:newdip} bottom 
indicate the level that might be achievable with future space
telescopes like {\it JWST}, if these are able to be sensitive to percent-level differences relative to the
photospheric flux.
This shows that such 12\,$\mu$m observations would be able to detect the brightening
expected from orbits with $q<0.3$\,au (see the brown line on the 12\,$\mu$m panel of
Fig.~\ref{fig:newdip} bottom).
That is, for a dip of this level, the non-detection of a change in near-IR flux at the sensitivity
of {\it WISE} means a detectable change should have occurred in the mid-IR (if we had 1\% photometric
precision).

\subsection{Implications for secular dimming constraints}
\label{ss:newdim}
A constant level of dimming is equivalent to the passage of a continuous sequence of dips.
Thus the total infrared flux, including the (extincted) stellar contribution,
can also be inferred from Fig.~\ref{fig:newdip}.
This is because the expected level of brightness increase should scale with the integral under
the different light curves in Fig.~\ref{fig:newdip} (since this is the integral of the flux
from material near pericentre, but with different pericentre passage times).
Rather than specify the necessary scaling, it can simply be noted that
for pericentres that are small enough for the contribution of stellar emission
to be ignored, the previous results from \S \ref{ss:irvt} apply.
Then it can be seen from Fig.~\ref{fig:newdip} that there is a pericentre distance 
for a given wavelength above which the increase in infrared flux from thermal
emission is more than balanced by the decrease in stellar emission;
e.g., this occurs at $q<0.05$\,au at 3.4\,$\mu$m, $q<0.1$\,au at 4.6\,$\mu$m and between
$0.1-0.3$\,au in the mid-IR.
Equivalently, for a given pericentre distance, there is a wavelength above which the
light curve changes from showing a decrease in emission to one in which an increase in
brightness is seen;
e.g., for $q=0.05$\,au this occurs at $\sim 3.4$\,$\mu$m and for $q=0.1$\,au this occurs
at $\sim 4.6$\,$\mu$m.

This has significant implications for the secular dimming, because \S \ref{sss:revised}
concluded that it is the 4.5\,$\mu$m upper limit which provides the strongest constraints
on the secular dimming, requiring $q>0.1$\,au to be consistent with 3\% dimming.
However, Fig.~\ref{fig:newdip} implies that an orbit with $q=0.1$\,au would exhibit very little
change in the observed (total) 4.5\,$\mu$m flux, irrespective of the level of dimming.
Practically this means that even lower pericentre distances are still compatible with the
secular dimming, down to $\sim 0.05$\,au.

\section{More accurate treatmenet of optically thick dust distributions}
\label{ss:thickedge}
A proper treatment of optically thick dust distributions would
require analysis using radiative transfer.
While codes for such analysis exist \citep[e.g.,][]{Dullemond2012}, the narrow eccentric ring geometry assumed here makes significant
modification necessary to allow resolution of the ring in a computationally efficient manner.
Thus far we have taken account of optical depth effects by calculating the absorbed radiation correctly
and then assuming this is reemitted isotropically with a spectrum that has the same shape as it would if
the dust distribution was optically thin.
Here we improve on this by acknowledging that the temperature of particles at different
locations will differ in an optically thick distribution (\S \ref{ss:thicktemp}), and by calculating the emission 
received for a specific viewing orientation based on the fraction of the ring that is not obscured by other parts
of the distribution (\S \ref{ss:thickview}).

\subsection{Dust temperature}
\label{ss:thicktemp}
\begin{figure}
  \begin{center}
    \vspace{-3.1in}
    \begin{tabular}{c}
      \hspace{-2.9in}
      \includegraphics[width=2.8\columnwidth]{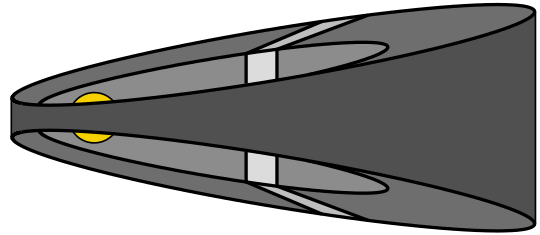}
    \end{tabular}
    \vspace{-7.5in}
    \caption{Edge-on view of the geometry shown in Fig.~\ref{fig:geom} left, noting again that the
    size of the star and the width and height of the distribution are exaggerated to illustrate the geometry.
    The shading shows the different faces of the distribution, with darkest shading for the outward
    face, then getting lighter for the top and bottom faces, and then lighter again for the starward face.
    The element shown on Fig.~\ref{fig:geom} left at an angle $\theta$ is shown with the lightest
    shading to illustrate the emitting area this element presents for this line of sight.
    The shading implies that the distribution is completely optically thick so that part of
    the emission from the starward face is obscured by ring material along the line of sight.
    }
   \label{fig:geom3}
  \end{center}
\end{figure}

Consider the angular element of the distribution which is shown by the darker shading in Fig.~\ref{fig:geom} left
at an angle $\theta \pm d\theta/2$ and distance $r \pm \Delta$.
The same element is shown on Fig.~\ref{fig:geom3} with the lighter shading assuming that the orbit is viewed exactly edge-on
(and $45^\circ$ from pericentre).
The width and height of the distribution is exagerrated on these figures, and for more realistic narrow distributions the
element cuts through an approximately square section of the distribution, with sides of length $2\kappa r$, and has a length
seen in projection from the star of $ds=rd\theta$ (see Fig.~\ref{fig:geom} left).

Stellar radiation is absorbed on only one face of the element which has an area $2\kappa r ds$, so that the
total energy absorbed is $E_{\rm abs}=[1-\exp{(-\langle\tau_{\rm abs}\rangle_\star)}]L_\star\kappa ds/(2\pi r)$.
We assume that there is no energy transfer to neighbouring elements and so this energy is deposited locally.
If the element contains a total cross-sectional area $d\sigma$, generalising equation~\ref{eq:taustar2} gives that
$\langle \tau_{\rm abs}\rangle_\star=\langle Q_{\rm abs} \rangle_{D,\star}d\sigma/(2\kappa r ds)$
(e.g., setting $d\sigma=\dot{\sigma}rds/h$ would recover something analogous to
equation~\ref{eq:tauextt}).
If the fraction of this area that is in particles of size $D$ to $D+dD$ is $\bar{\sigma}(D)dD$, then particles
in this size range in the element absorb an amount of energy
$E_{\rm abs}\bar{\sigma}(D)dD\langle Q_{\rm abs}\rangle_\star/\langle Q_{\rm abs} \rangle_{D,\star}$.

The dust temperature is calculated by balancing the absorbed radiation with that reemitted,
which for particles in the range $D$ to $D+dD$ is
$E_{\rm em}=S \bar{\sigma}(D)dD\int Q_{\rm abs}(\lambda,D)B_\nu(\lambda,T)d\lambda$.
Here $S$ is the total emitting area of the element.
For optically thin distributions this is simply $4d\sigma$, however this cannot increase indefinitely with $d\sigma$.
We consider two assumptions for the emitting area in the optically thick case. 
If the absorbed energy is reradiated equally from the four external faces of the element then $S$ can only
increase up to a maximum of $8\kappa rdsJ(\theta)$, where
\begin{equation}
  J(\theta) = \sqrt{1+e^2+2e\cos{\theta}}/(1+e\cos{\theta})
  \label{eq:j}
\end{equation}
is a factor that accounts for the fact that the length of the element along the orbit is longer than its projected length $ds$;
this factor is unity at pericentre and apocentre, and has a maximum of $(1-e^2)^{-1/2}$ at the true anomaly
for which $r=a$.
Alternatively, if the absorbed energy is only reradiated from the starward face of the element then $S$
can only increase up to $2\kappa rdsJ(\theta)$.

\begin{figure*}
  \begin{center}
    \begin{tabular}{cc}
      \includegraphics[width=1\columnwidth]{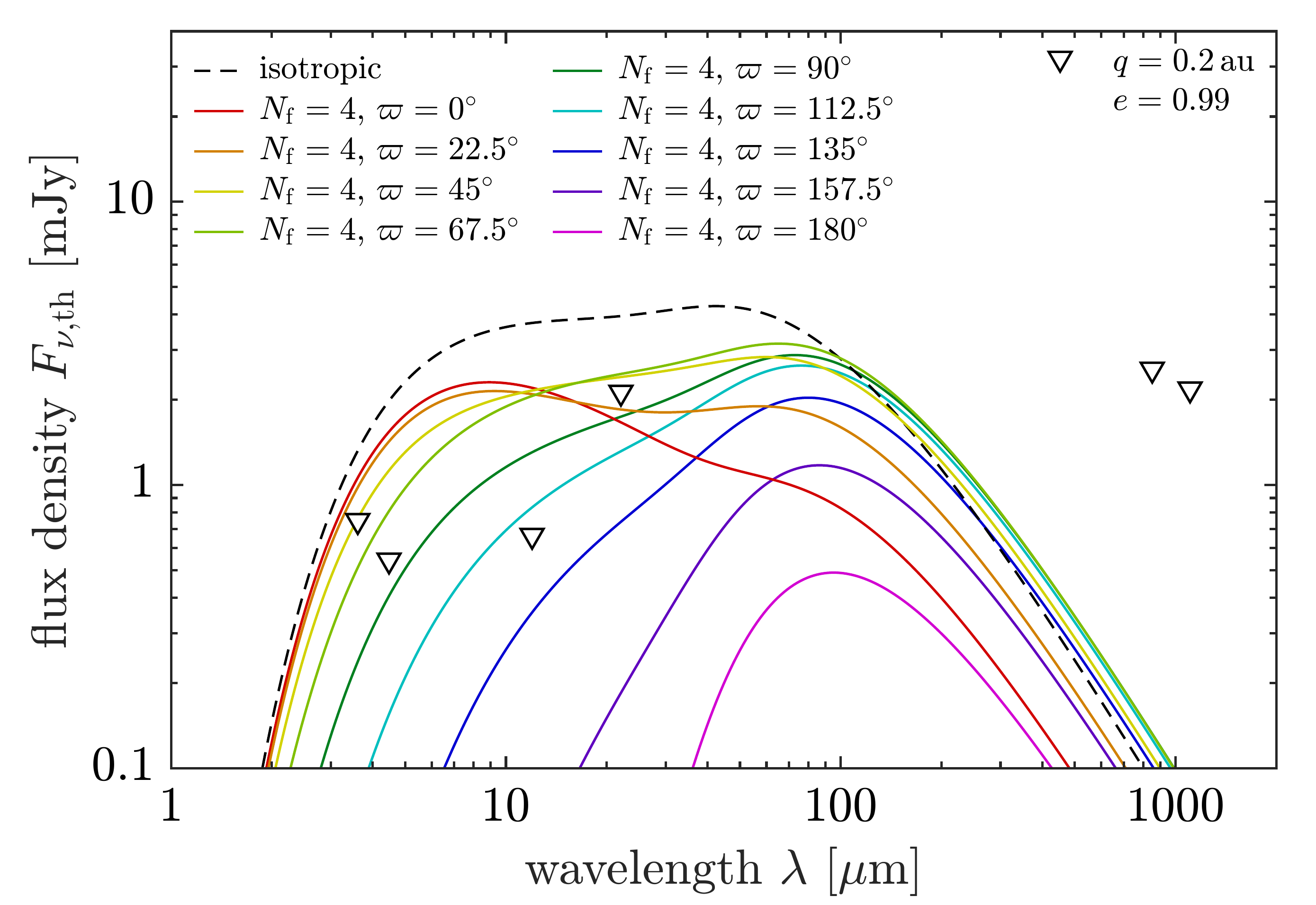} &
      \includegraphics[width=1\columnwidth]{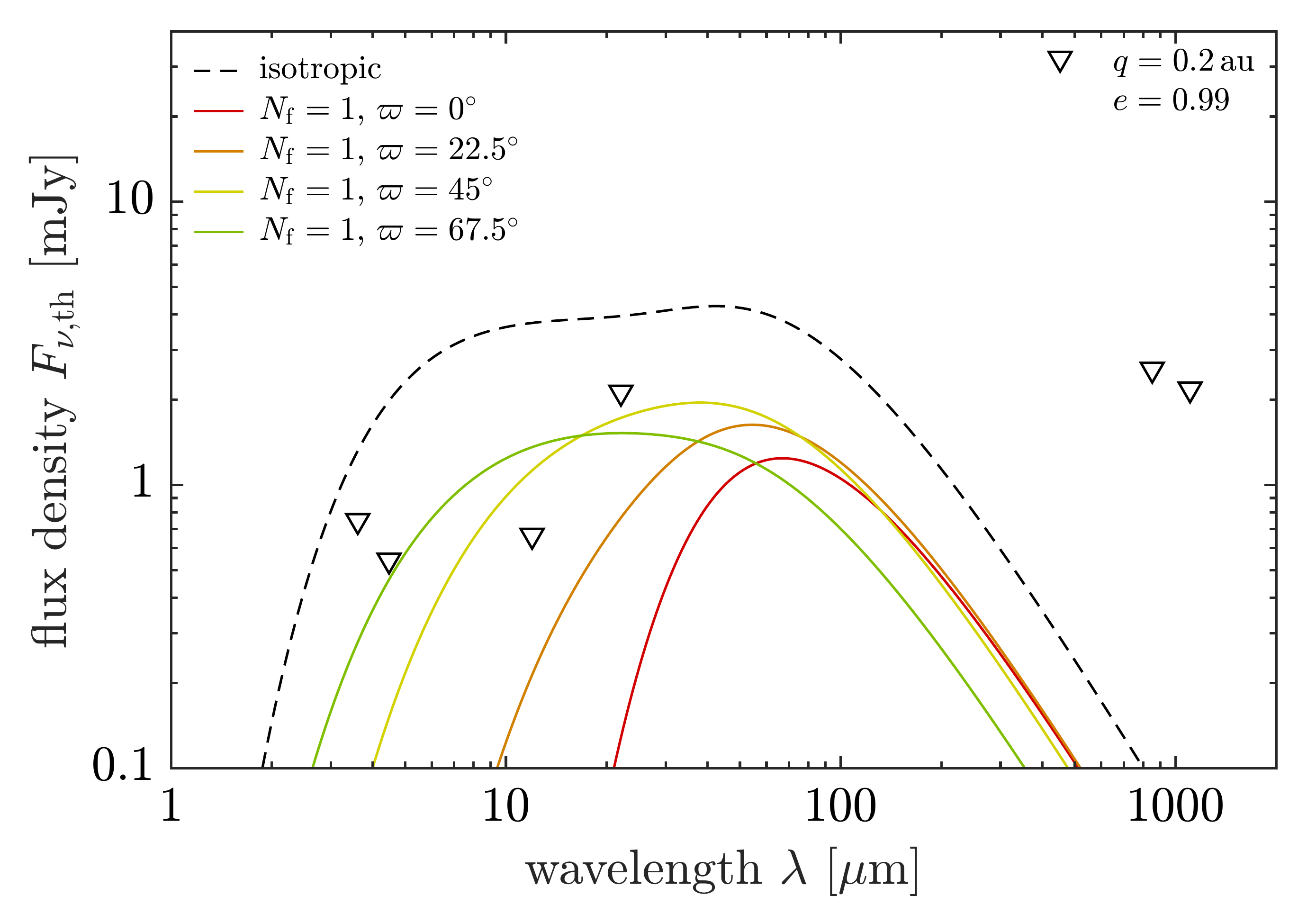} \\
    \end{tabular}
    \caption{
    Spectrum of thermal emission for an optically thick edge-on ($i=0$) black body
    dust distribution, for an orbit with $q=0.2$\,au and $e=0.99$ with a vertical scale height 
    sufficient to cause 10\% dimming when observed along pericentre
    (i.e., $\kappa=0.1(\pi/4)(R_\star/q)\approx 3 \times 10^{-3}$, see eq.~\ref{eq:omst}).
    The dashed line shows the predicted spectrum calculated with the assumption that the absorbed
    light is emitted isotopically and at the same temperature as for optically thin dust distributions,
    where black body dust properties were assumed.
    The solid lines show the more realistic case for optically thick distributions where the
    emission depends on viewing orientation, which is both edge-on and has different
    pericentre orientations for different coloured lines as shown in the legend.
    The level of dimming depends on pericentre orientation, increasing from 10\% by a factor
    that can be derived from equations~\ref{eq:rt} and \ref{eq:omstar}.
    The left panel assumes that the absorbed radiation is emitted over all surfaces of
    distribution ($N_{\rm f}=4$), which is valid for marginally optically thick distributions.
    The right panel assumes that the absorbed radiation is only emitted from the
    starward face ($N_{\rm f}=1$), which is more accurate for very optically thick distributions.
    }
   \label{fig:thick}
  \end{center}
\end{figure*}

The balance of absorbed and emitted energy then gives the dust temperature to be the same as equation~\ref{eq:tdr},
but with an additional factor of $\chi$ on the right hand side, where
\begin{equation}
  \chi^4  = \frac{ 1-\exp{(-\langle \tau_{\rm abs}\rangle_\star)} } 
              { {\rm min}[~\langle \tau_{\rm abs}\rangle_\star~~,~~(N_{\rm f}/4) \langle Q_{\rm{abs}} \rangle_{D,\star} J(\theta)~]},
          \label{eq:tdr2}
\end{equation}
where the denominator takes whichever is smaller of the two values, and $N_{\rm f}$ is the number of faces over which radiation is
emitted (i.e., $N_{\rm f}=1$ or 4 depending on whether emission is only over the starward face or over all four external
faces).
For optically thin distributions $\chi=1$ as expected.
For optically thick distributions that emit over all four external faces
$\chi=[(J(\theta)\langle Q_{\rm{abs}} \rangle_{D,\star}]^{-1/4}$,
which is also close to unity at pericentre and apocentre, and lower in between.
For optically thick distributions that only emit from the starward face
$\chi=\sqrt{2}[(J(\theta)\langle Q_{\rm{abs}} \rangle_{D,\star}]^{-1/4}$,
which is hotter than the optically thin case at pericentre and apocentre, but can be
lower in between for large eccentricities.

In reality the temperature will be neither uniform
across all external faces, nor will it be zero on those faces not pointing toward the star.
However, these two assumptions provide bounding cases that may be reasonable
approximations when the distribution is marginally optically thick (emission over all four faces)
or highly optically thick (emission over just the starward face).

\subsection{Emitting area}
\label{ss:thickview}
To calculate the observed emission in equations~\ref{eq:fnu} and \ref{eq:fnut}, it was assumed that all of a particle's
cross-sectional area is visible along our line of sight. 
While this is true for optically thin distributions, the same considerations that applied in \S \ref{ss:thicktemp}
regarding the area over which the absorbed energy is reemitted must also apply here, otherwise energy
would not be conserved.
In this case we must also account for the fact that the face is not necessarily normal
to our line of sight;
for a face $j$ with a normal (that faces into the element) that makes an angle $\phi_j$ to our line of sight,
its emitting area is $2\kappa r^2 d\theta J(\theta) \cos{\phi_j}$, and is hidden from view if $\phi_j>90^\circ$.
For a viewing geometry which is perfectly edge-on ($i=0$), the back face has
$\cos{\phi_j}=[\cos{(\theta+\varpi)}+e\cos{\varpi}]/\sqrt{1+e^2+2e\cos{\theta}}$, while the starward face
has a minus sign in front of this expression, and the top and bottom faces have an extra factor of
$\kappa$ on the right hand side (assuming $\kappa \ll 1$).
This reduction in area is clearly seen in Fig.~\ref{fig:geom3}, which also shows that a further reduction
in area may be needed when another portion of the distribution lies in front
of the element (which applies to the starward face in this figure).
This reduction is a purely geometrical factor, which we call $\eta_j$, though it should also include 
consideration of the extinction at wavelengths appropriate for the emitted radiation which is at 
longer wavelengths than the stellar emission (and so extinction is less efficient, see Fig.~\ref{fig:dldv}).

Combining these reduction factors, we define the fraction of face $j$ that is visible to our line of sight as
\begin{equation}
  K_j = {\rm max}[\cos{\phi_j}~,~0]\eta_j.
\end{equation}
Thus the total emitting area of the element is
\begin{equation}
  dA = 2\kappa r^2 d\theta J(\theta) \sum_j K_j,
  \label{eq:da}
\end{equation}
where the sum should be over the four external faces, or just include the starward face,
depending on the assumption used for the temperature calculation.
This can then be included in equations~\ref{eq:fnu} and \ref{eq:fnut} as a constraint on the 
maximum possible area from the element.
This can be achieved in equation~\ref{eq:fnu} by replacing
$d\sigma/d\theta$ with $dA/d\theta$ (from eq.~\ref{eq:da})
if $\langle\tau_{\rm abs}\rangle_\star > (N_{\rm f}/4) \langle Q_{\rm{abs}} \rangle_{D,\star} J(\theta)$.
This change arises when the emitting area is set by the surface area of the element (in which case the emission is not
isotropic), rather than by the total surface area of dust particles (in which case the emission is isotropic),
and occurs at the same optical depth as the change in the denominator in eq.~\ref{eq:tdr2}.
Note that the $[1-\exp{(-\langle\tau_{\rm abs}\rangle_\star)}]/\langle\tau_{\rm abs}\rangle_\star$ factor
that was added to eq.~\ref{eq:fnu} in \S \ref{ss:irobs} is no longer needed because it is included in the temperature calculation,
which must now include the $\chi$ factor from eq.~\ref{eq:tdr2}.
For equation~\ref{eq:fnut} the required change is to replace
$\dot{\sigma}(t')$ with $2\kappa h J(\theta) \sum_j K_j$ when $\dot{\sigma}(t')>2\kappa h J(\theta) (N_{\rm f}/4)$
(where the choice was translated from a limit in optical depth to one in the rate at which cross-sectional
area crosses the star using eq.~\ref{eq:tauextt}).

Figure~\ref{fig:thick} shows the effect of the above considerations for the emission spectrum of an optically
thick distribution viewed edge-on {\bf (the orientation with the greatest reduction in near- and mid-IR emission)},
for material evenly distributed around an orbit with $q=0.2$\,au and $e=0.99$.
Black body dust is assumed so that it is easy to see the contribution to the spectrum from dust at pericentre
and apocentre (see discussion of Fig.~\ref{fig:spec} in \S \ref{ss:irobs}).
The spectrum that would have been inferred by assuming the absorbed radiation is emitted isotropically at temperatures
corresponding to those of optically thin dust is shown with a dashed line. 
The left panel of Fig.~\ref{fig:thick} shows how the spectrum is modified with the more accurate treatment
of optical depth effects when each element is assumed to emit the radiation across all four faces for
different pericentre orientations.
There is a clear overall reduction in emission due to the lower cross-sectional area of the distribution that is presented
to the observer (i.e., because more of the absorbed radiation is emitted perpendicular to the orbital plane).
The temperature of the emission is also modified slightly.
Self absorption of the emission is also evident for viewing orientations close to apocentre, since for this
orientation (remembering that we are also assuming a completely edge-on orientation)
the pericentre is completely obscured by the optically thick material at apocentre resulting
in the removal of the hot near-IR and mid-IR emission.
For viewing orientations close to pericentre, the resulting change to the spectrum is more modest, particularly
at near-IR wavelengths, but is reduced by up to a factor 4 at longer wavelengths.
The right panel of Fig.~\ref{fig:thick} shows the same as the left panel, but this time assuming that each element
only emits radiation from its starward face.
Again the overall level of emission is noticeably reduced, but with hot near-IR emission from material near pericentre
conspicuously absent for all orientations (since it is always obscured).
No emission is seen at all for orientations with $\varpi>90^\circ$.
Thus optical depth effects can have a drastic effect on the level of predicted emission.

\subsection{Implications for secular dimming constraints}
\label{ss:finalimplic}
The implication of Fig.~\ref{fig:thick} for the constraints that the lack of infrared emission place on
the secular dimming are considerable. 
Until now we have concluded that the infrared observations require the material to pass the pericentre
at a distance of at least 0.05\,au.
This is no longer necessary if the distribution is optically thick, since it is possible to hide the
hot near-IR emission that provides the most stringent constraints by self-absorption.
That is, all of the stellar radiation that is absorbed at pericentre can be reemitted in
directions other than towards an observer oriented toward pericentre
(which also means that if the distribution is optically thick then non-transiting KIC8462852-like systems that 
are more face-on would be more readily detectable in the infrared).

In fact, the same is true of the mid-IR emission which can also be completely blocked by self-absorption
(see Fig.~\ref{fig:thick}).
Thus the conclusion that the century-long dimming had to have gone away by the time {\it Kepler} started observing
(i.e., before the {\it WISE} 12\,$\mu$m measurement in 2010 May) is also no longer required, assuming that the
dust distribution is optically thick enough for elements to emit only from their starward faces.
Thus it appears that there are no strong constraints from the lack of infrared emission on the
orbit, and all levels of proposed secular dimming are compatible with dimming by circumstellar material.

\begin{figure}
  \begin{center}
    \begin{tabular}{c}
      \includegraphics[width=1\columnwidth]{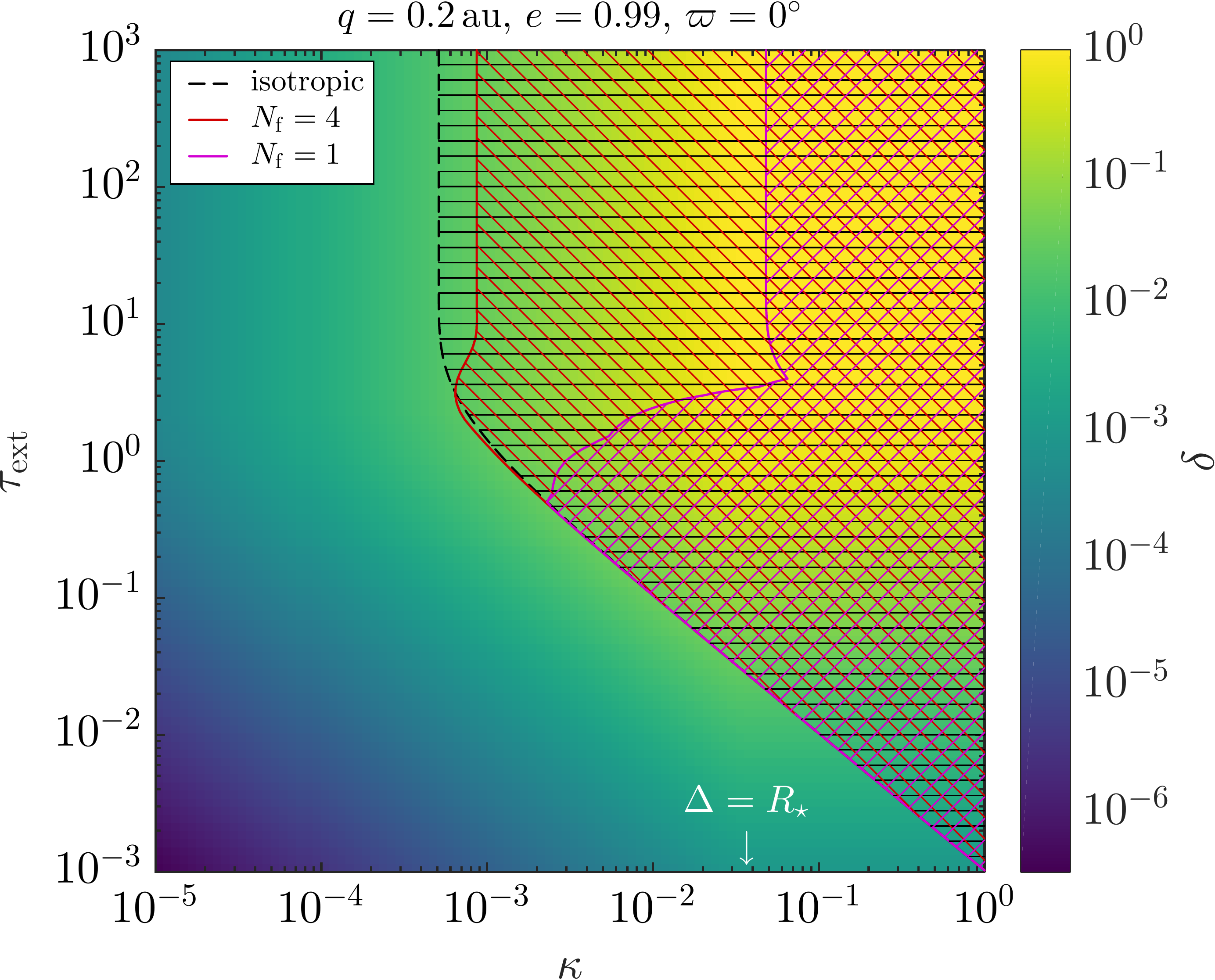}
    \end{tabular}
    \caption{
    The effect of different treatments for optical depth effects on the maximum level of dimming
    allowed by the infrared constraints.
    An orbit with $q=0.2$\,au and $e=0.99$ is assumed to be viewed toward pericentre ($\varpi=0^\circ$).
    The colour scale shows the level of dimming expected for a given optical depth $\tau_{\rm ext}$
    and vertical scale height $\kappa$ for the distribution.
    The cross-hatched regions show the region of parameter space excluded by the infrared observations
    assuming black body dust.
    The dashed line and horizontal hatching assumes the emission is isotropic and at the same temperature
    as optically thin dust, while the red and purple solid lines and diagonal hatching makes the assumptions
    of \S \ref{ss:thickedge} for elements that emit over all external surfaces ($N_{\rm f}=4$) and
    those that only emit over the starward face ($N_{\rm f}=1$).
    }
   \label{fig:tauvk}
  \end{center}
\end{figure}

To illustrate this more quantitatively, consider Fig.~\ref{fig:tauvk} which shows for a specific orbit
($q=0.2$\,au and $e=0.99$) and orientation ($\varpi=0^\circ$) the level of dimming expected for
distributions with different optical depths $\tau_{\rm ext}$ and vertical scale heights $\kappa$.
The dimming levels come from equations~\ref{eq:omstar} and \ref{eq:deltadef}.
The lines and cross-hatched regions show the areas of parameter space excluded by
the infrared observations with different treatments of optical depth effects.
The treatment of \S \ref{ss:irobs} in which optically thick emission is assumed to be
isotropic at the temperature expected for optically thin dust is shown with a dashed line
and horizontal hatching.
For this treatment the level of dimming in the allowed parameter space has a maximum value
which is the level plotted in Fig.~\ref{fig:ir1}.
The treatments of \S \ref{ss:thickview} with $N_{\rm f}=4$ and $N_{\rm f}=1$ are shown with red and
purple lines, respectively. 
These show that optical depth effects only make a small change if the distribution can emit across
all faces of the element, but that the region of allowed parameter space is much larger when 
the emission is only from the starward face of the element.
In particular, for large optical depths the infrared upper limits can accommodate distributions
(for this configuration) that are vertically broad enough to completely cover the star and so
result in a dimming $\delta=1$.

\section{Transit probability}
\label{s:mc}
The probability of witnessing material transit in front of the star at a distance of $r_{\rm t}$ is
\begin{equation}
  p = \kappa + R_\star/r_{\rm t},
  \label{eq:p}
\end{equation}
where the term involving $\kappa$ accounts for the finite height of the distribution about the orbit.
This provides the basis for consideration of the pericentre orientation expected for detections of
phenomena like that seen toward KIC8462852 (\S \ref{ss:periapo}), and the probability of witnessing
such phenomena (\S \ref{ss:rarity}), as well as to consider what future transit surveys might detect
(\S \ref{ss:pop}).

\subsection{Pericentre vs apocentre transits}
\label{ss:periapo}
Equation~\ref{eq:p} means that, for a given orbit, we are more likely to see a transit close to pericentre
than close to apocentre;
e.g., the fraction of transits seen with pericentre orientations in the range 0 to $\varpi$ is
$[\varpi + e\sin{\varpi}]/\pi$.
However, another consideration is that for the assumption of a narrow distribution
in eq.~\ref{eq:omst}, equation~\ref{eq:deltadef} shows that, for the same amount of material
in the same orbit, the dimming is greater at apocentre than at pericentre by a factor $(1+e)/(1-e)$.
Thus if there is a dimming threshold below which we would not have detected this phenomenon, the first
detection might be expected close to apocentre.
However, if the vertical extent of the distribution is such that it extends beyond the
line of sight to the star at apocentre, the assumption of eq.~\ref{eq:omst} breaks down.
If this is the case, transits at all radii beyond the distance at which the distribution fully covers the
star have the same depth, but those at larger radii have a lower probability of being observed.

Applying these qualitative arguments to KIC8462852, consider that if we are observing this phenomenon
at pericentre then we might expect that other hypothetical observers (on other stars doing analogous {\it Kepler} surveys)
see an even greater dip depth or dimming.
However, they would on average need to survey a larger number of stars than the 150,000 surveyed by {\it Kepler}
to see this phenomenon. 
Similarly, if we undertake a larger survey (like PLATO) we might expect to see some stars with greater
levels of dips/dimming than KIC8462852.
Conversely, if we are observing this phenomenon at apocentre then this phenomenon would be seen more
frequently by other observers (that are aligned closer to the pericentre), but with smaller dips/dimming;
i.e., if we see one dip at apocentre from KIC8462852 we would expect many other stars with the
same dust structures that exhibit lower level dips from transits at pericentre.  
However, the considerations in \S \ref{ss:shortdip} already rule out an apocentre orientation for the
transit for KIC8462852 using considerations which are not affected by the subsequent discussion
in \S \ref{s:dist}, \ref{s:stellardim} and \ref{ss:thickedge} (e.g., this can be inferred from the presence
of a 0.4\,day dip and the requirement for a period longer than 750\,days).

\subsection{Rarity of KIC8462852-like dust distributions}
\label{ss:rarity}
Equation~\ref{eq:rmin} showed that the lack of infrared emission requires a minimum distance at which the dust
must transit the star.
While we later discussed ways in which this constraint can be relaxed (\S \ref{ss:irobs}, \ref{ss:shortdip},
\ref{s:dist}, \ref{s:stellardim} and \ref{ss:thickedge}),
to allow it to accommodate the fact that the shortest dips require a maximum distance at which the
dust transits the star (equation~\ref{eq:rminsd}), it is inevitable that the lack of observed infrared emission
means that there is a minimum distance at which transits occur.
Unless the dust distribution is completely opaque, this minimum distance is 0.05\,au.
This has further implications through eq.~\ref{eq:p} for the probability of witnessing a system like this.

It is dangerous to consider statistics on a sample of 1, since we cannot exclude that the phenomenon is
rare and we just had a favourable orientation for this system. 
However, if we take the combined constraints to imply a transit distance of $\sim 0.1$\,au, this implies that there could
be $\sim 14$ stars in the {\it Kepler} sample with a similar phenomenon but oriented unfavourably to see the material
transit the star.
This would mean that approximately 1 in $10^4$ stars exhibit this phenomenon (i.e., with detectable quantities of
dust distributed around an eccentric orbit) in any given time interval corresponding to the {\it Kepler} lifetime of
$\sim 1500$\,days.
This could imply that all stars have such a dusty eccentric orbit for $\sim 0.5$\,Myr of their lives,
or 0.01\% of stars have such (mega-)structures throughout their lives.

\subsection{Distribution of dimming in population model}
\label{ss:pop}
While \S \ref{ss:rarity} discussed the rarity of the KIC8462852 phenomenon, this cannot be properly
assessed from just one detection.
Thus, this section considers the prospects for detecting transits from circumstellar material around other stars
in future surveys.

In \S \ref{ss:periapo} we noted that for a given dust mass on a given orbit 
the orientation of the pericentre of the orbit to our line of sight affects both
the probability of witnessing a transit and the depth of dimming we might witness.
Here we expand on this to consider a Monte Carlo model for the distribution of dimming events
we might expect to see in a population of stars.
We assume that all stars have the same radius $R_\star$, and have debris
distributed around the same orbit, which is characterized by its pericentre and eccentricity $q$ and $e$,
as well as the height of the distribution about that orbit $\kappa$.
The total amount of cross-sectional area in the debris is taken from a power law distribution
in the range $\sigma_{\rm min}$ to $\sigma_{\rm max}$, with $n(\sigma)d\sigma \propto \sigma^{\gamma}$.
The inclination of the orbit to our line of sight $i$ is assumed to be randomly distributed
in $\sin{i}$.
The orientation of the pericentre with respect to the line of sight $\varpi$ is
also assumed to be random, though we use the model to highlight the difference between
observing the transit at pericentre ($\varpi=0$) and apocentre ($\varpi=\pi$).

\begin{figure}
  \begin{center}
    \vspace{-0.5in}
    \begin{tabular}{c}
      \hspace{-0.5in}
      \includegraphics[width=1.2\columnwidth]{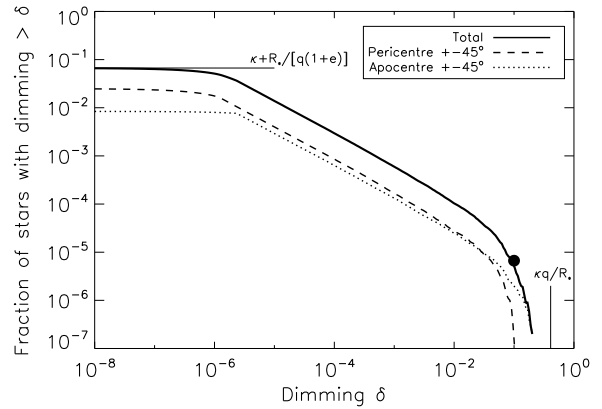}
    \end{tabular}
    \vspace{-2.2in}
    \caption{Fraction of stars with dimming above a given $\delta$ for a model population in which
    all stars have material evenly distributed around orbits with $q=0.1$\,au, $e=0.9975$,
    $\kappa=0.03$, with cross-sectional area drawn from a distribution with $n(\sigma) \propto \sigma^{-5/3}$
    between $\sigma_{\rm min}=10^{-4}$\,au$^{2}$ and $\sigma_{\rm max}=10$\,au$^{2}$.
    The black circle shows the point that the distribution would pass near to expect to
    see one example of KIC8462852 in the {\it Kepler} sample.
    }
   \label{fig:ngtd}
  \end{center}
\end{figure}

Fig.~\ref{fig:ngtd} shows the distribution of dimming levels for a population with parameters given
in the caption (calculated using equations~\ref{eq:omstar}, \ref{eq:deltadef}, \ref{eq:f} and \ref{eq:tauext}).
It would be possible to get constraints on the model parameters that result in a distribution that goes
through the expectation of one KIC8462852-like object in the {\it Kepler} survey.
For example, equation~\ref{eq:omst} shows that the maximum possible dimming for pericentre transits is
$\sim \kappa q/R_\star$, and averaging equation~\ref{eq:p} over all possible pericentre orientations
using equation~\ref{eq:rt} shows that the fraction of stars with dimming at any level is
$\kappa+R_\star/[q(1+e)]$.
While Fig.~\ref{fig:ngtd} shows that both of these limits provide a useful guide to the boundaries of the
resulting distribution, going further with such an analysis would place undue emphasis on the model.
Consider, for example, that we could have made many
assumptions about the distribution of cross-sectional area or orbital properties in the population.
Rather we show this simply to illustrate that it is possible to construct a population model
wherein dimming at the level seen toward KIC8462852 is expected.

This example population also shows how the detection of one object with a large level of dimming
implies the existence of many more with lower levels of dimming, with analogous implications for
the detection of one object with large dips;
e.g., the population model of Fig.~\ref{fig:ngtd} has 0.1\% dimming toward 1:7,700 stars
(i.e., suggesting that $\sim 20$ of the {\it Kepler} stars should exhibit 0.1\% dimming or dips), 
and 1\% dimming toward 1:35,000 stars (i.e., suggesting that $\sim 4$ of the {\it Kepler} stars should
exhibit dimming or dips at this level).
These would be the same system viewed with more or less favourable viewing geometries, and
those with lower quantities of dust.
Future transit surveys such as PLATO \citep{Rauer2014} may detect many more systems with transiting
circumstellar material that would constrain the distribution of dimming around nearby stars, and
so could eventually be used to determine the frequency and orbits of exocomet-like dust
distributions.

Comparing the number of stars with dimming detected near pericentre (dashed line on Fig.~\ref{fig:ngtd})
with those detected at the same dimming levels near apocentre (dotted line),
we find that large dimming levels are expected to be detected at a similar rate for the two orientations.
While this may appear surprising given the lower probability for transits at apocentre, it arises
naturally because of the larger dimming level for apocentre orientations as already noted.
However, it is worth noting that for KIC8462852 its secular dimming was only identified because of the
short term dimming events.
It is likely that short term dimming events are easier to identify when viewed along pericentre, because
such events last longer by a factor $(1+e)/(1-e)$ when viewed at apocentre than at pericentre (e.g.,
turning a 0.4\,day event into one of duration 320\,days for $e=0.9975$).
It could also be that the dust level is enhanced near pericentre due to cometary activity making
it more readily identifiable than the model here would predict.
If this were the case we might expect the short term dimming events to be dominated by smaller
dust than the secular dimming, which would result in a different wavelength dependence for the dimming
(depending on whether the material is optically thin, e.g. see Fig.~\ref{fig:dldv}).

\section{Conclusion}
\label{s:conc}
Here we have developed a model for the interpretation of a star's optical light curve under the assumption that
any dimming is caused by circumstellar material spread around a single orbit (i.e., a more general form
of {\it exocomet} model).
In particular the model is used to predict the level of infrared flux from thermal emission
from the material, as well as the wavelength dependence of the dimming of the starlight.
The model is focussed on application to the light curve of KIC8462852 which shows both secular dimming
and short duration dips.

At first in \S \ref{s:ir} the dust was assumed to be evenly distributed around the orbit, which is appropriate
for an approximately constant level of secular dimming.
This showed that:
\vspace{-0.7em}
\begin{itemize}
\item
The fractional luminosity of the thermal emission is related to the
level of dimming by a factor that depends only on the distance of the material from the
star at the point of transit (see \S \ref{ss:flim} and eq.~\ref{eq:delta2}).
\item
Considering the upper limits on the thermal emission at different wavelengths (but not the
timing of those measurements), the 12\,$\mu$m {\it WISE} observation sets the strongest
constraints on the orbit for the infrared observations to be consistent with the observed
level of secular dimming (see \S \ref{ss:irobs} and Fig.~\ref{fig:ir1}).
\item
The shortest 0.4\,day dip in the {\it Kepler} light curve requires the material to transit
the star at $<0.6$\,au.
\item
The preliminary conclusion of \S \ref{ss:shortdip} was that the 12\,$\mu$m {\it WISE} observation
ruled out that the putative century long 16\% dimming was still present at the epoch of {\it Kepler}
(assuming that this dimming originates in circumstellar material on the same orbit as that
causing the short duration dips), and required that
the orbit has a pericentre in the $0.1-0.6$\,au range, and is observed within $\sim 90^\circ$
of that pericentre.
\end{itemize}

There are, however, three important caveats to this conclusion that required further development
of the model: the dimming is time variable, the observed infrared flux includes the contribution
from the stellar flux in addition to the dust thermal emission, and optically thick distributions
had only been accounted for in an approximate way.

The time variability of the dimming was accounted for in \S \ref{s:dist} by developing a method to use the optical light
curve to derive the uneven distribution of dust around the orbit and so predict the level of
thermal emission expected at the epochs of the infrared measurements (see Fig.~\ref{fig:irvt}).
This showed that:
\vspace{-0.7em}
\begin{itemize}
\item
Since the 12\,$\mu$m measurement was made before the most securely determined
secular dimming (i.e., that measured by {\it Kepler}), the 4.5\,$\mu$m measurements were instead 
those most constraining (see \S \ref{sss:revised}).
\item
It was still concluded that the century-long secular dimming was likely absent by the time
{\it Kepler} started observing
(unless the secular dimming does not originate in circumstellar material on the same orbit as that causing
the short duration dips), and the constraints on the orbit still required an orbit that transits
the star beyond 0.1\,au (and within 0.6\,au).
\item
The thermal emission is expected to undergo epochs of brightening as the clumps of material
associated with the dips pass through pericentre (see \S \ref{sss:spike}).
\item
The infrared brightening can be slightly offset from the optical dips if the transit is not oriented
exactly along pericentre, and lasts 10s of days at near- and mid-IR wavelengths.
\item
The peak near-IR flux from the thermal emission brightening depends on the integrated dip depth, as well as the pericentre
distance, and can approach levels that are detectable with current instrumentation (see Fig.~\ref{fig:irvq}).
\end{itemize}

In \S \ref{s:stellardim} it was shown to be important to include the stellar contribution when considering near- and mid-IR
constraints, because the wavelength dependence of the dimming is not known a priori.
This contribution was included in the analysis and used in \S \ref{ss:newdip} to predict the light curves at different
wavelengths for a 2\% dip that is similar in character to one recently observed in the KIC8462852 light curve (see Fig.~\ref{fig:newdip}),
and was also applied to the secular dimming constraints in \S \ref{ss:newdim}.
It was concluded that:
\vspace{-0.7em}
\begin{itemize}
\item
For the assumed dust optical properties and a size distribution with a minimum size 
set by radiation pressure, dimming at optical wavelengths is independent of wavelength (see Fig.~\ref{fig:dldv}), consistent
with recent measurements of the secular dimming.
\item
A smaller minimum grain size would result in wavelength dependent extinction, but would still result in a
grey extinction if the dust distribution is optically thick.
\item
The stellar contribution to the observed flux from a 2\% dip can only be ignored for pericentres that are very close
to the star, or for long wavelengths.
\item
Infrared observations during dipping events with {\it Spitzer} or {\it JWST} could detect the predicted changes in the level of 
infrared emission (and so be used to constrain the orbit, pericentre orientation and optical depth).
\item
For orbits with a pericentre at 0.1\,au, an azimuthally broad distribution of material that causes secular dimming 
would result in a decrease in total flux at wavelengths shorter than $\sim 5$\,$\mu$m, and
an increase in flux at longer wavelengths.
\item
This results in a decrease in the minimum pericentre distance allowed for the secular
dimming to be consistent with the infrared constraints, so that now orbits that transit the
star $0.05-0.6$\,au are allowed.
\end{itemize}

\begin{table}
  \centering
  \caption{Conclusions on the distance of material from the star at the point of transit ($r_{\rm t}$)
  for different assumptions about the optical depth of the dust distribution and about the level of secular
  dimming that was present at the start of the {\it Kepler} observations.}
  \label{tab:sum}
  \begin{tabular}{l|cc}
     \hline
     & \multicolumn{2}{c}{Secular dimming in 2010 May} \\
     & 16\% & 0\% \\
     \hline
     Optically thin  & Not allowed &  $0.05$\,au$<r_{\rm t}<0.6$\,au \\
     Optically thick & $r_{\rm t}<0.6$\,au & $r_{\rm t}<0.6$\,au \\
     \hline
  \end{tabular}
\end{table}

Thus far optical depth effects had been accounted for, but on the assumption that the
absorbed radiation is emitted isotropically at the temperature expected for optically thin dust.
To improve on this in \S \ref{ss:thickedge} we developed the analytical calculation to consider highly optically thick
distributions for which all of the absorbed radiation is emitted from the side of the distribution
facing the star, and marginally optically thick distributions for which the absorbed radiation is
emitted equally from all surfaces of the distribution.
This showed that:
\vspace{-0.7em}
\begin{itemize}
\item
Optical depth effects can easily hide hot emission from dust at pericentre by self-absorption.
\item
All of the constraints on the orbit from the infrared emission can be
circumvented for highly optically thick distributions (see \S \ref{ss:finalimplic});
that is, either the material transits the star at a distance of $0.05-0.6$\,au, or it is 
highly optically thick in which case the only constraint is that it must transit $<0.6$\,au.
\item
Highly optically thick distributions would also allow the century-long 16\% secular dimming
to be present at the epoch of the {\it Kepler} observations.
\item
Thus the infrared constraints are compatible with a circumstellar origin (e.g., exocomets) for the
KIC8462852 light curve, with constraints on the orbit that are summarised in Table~\ref{tab:sum}.
\end{itemize}

Finally, we considered in \S \ref{s:mc} the probability of witnessing such a phenomenon in a population of
stars all with material evenly distributed around elliptical orbits. 
This showed that:
\vspace{-0.7em}
\begin{itemize}
\item
It is possible to construct a population model in which the detection of
one KIC8462852-like system in the {\it Kepler} sample is a reasonable expectation.
\item
Perhaps surprisingly, large levels of dimming may be associated with viewing
orientations that see the orbit near apocentre, although dips may be more readily detectable
for pericentre-oriented lines of sight.
\item
Lower levels of dimming (and dips) should be more common and
so we can expect further detections of this phenomenon with future large transit surveys like
PLATO.
\end{itemize}

To conclude, we find that the short and long term variability in the optical light curve of KIC8462852,
and the lack of variability at current detection thresholds in the infrared light curves, are
compatible with an origin in circumstellar material distributed around an elliptical orbit.
Constraints were placed on that orbit and the uneven distribution of dust around the
orbit derived, and we showed that there are good prospects for detecting variability in the infrared light curve
at wavelengths beyond 3\,$\mu$m with, for example, {\it JWST}.
However, it is important to remember that this model was focussed on showing what is needed to satisfy
the observational constraints, and made no consideration of the origin of this elliptical dust structure.
The inferred highly elliptical orbit and analogy with sungrazing comets in the Solar System \citep[e.g.,][]{Sekanina2004}
suggests a scenario in which one massive exocomet fragmented into multiple bodies which, due to their slightly different
orbits, are now at different longitudes but close to the orbit of the original parent body.
These large fragments continue to fragment and release dust thus replenishing the observed dust structure.
In this scenario the largest gravitationally bound dust ends up distributed around the orbit due to Keplerian shear and is
responsible for the secular dimming, while dust recently released from the fragments is responsible for the short duration
dips.
However, this scenario needs to be further explored using modelling like that of \citet{Bodman2016}
and \citet{Neslusan2017} to assess whether this is compatible with the detailed shapes of the dips, as well
as whether the required total mass and fragmentation history are realistic.
Searches for the parent belt from which the exocomet originated are also warranted (e.g., using sub-mm observations
with {\it ALMA}), as are studies to consider the dynamical origin of the exocomet which could, e.g., be linked to the 
nearby companion star \citep{Boyajian2016}.
Other scenarios may also be able to explain the light curves of KIC8462852, but any scenario that invokes
circumstellar material \citep[e.g.,][]{Ballesteros2017} would have to satisfy the constraints discussed in this paper.

\section*{Acknowledgments}
GMK is supported by the Royal Society as a Royal Society University Research Fellow.

\appendix

\section{Parameter summary}

\begin{table*}
  \centering
  \caption{Summary of the parameters used in the paper.}
  \label{tab:sym}
  \begin{tabular}{llll}
     \hline
     Parameter               & Meaning \\
     \hline
     $dA$                    & Emitting area of an angular element of the distribution as seen along our line of sight \\ 
     $b$                     & Impact parameter; projected distance above the centre of the star \\
     $B_\nu(\lambda,T)$      & Planck function at wavelength $\lambda$ for temperature $T$ \\
     $d$                     & Distance to star from Earth \\
     $D$                     & Particle diameter \\
     $e$                     & Eccentricity of orbit \\
     $f$                     & Fractional luminosity of material \\
     $f_{\rm BB}$            & Fractional luminosity material would have if the dust behaved like a black body \\
     $F_{\nu,{\rm obs}}$     & Total flux density observed, including dust thermal emission and stellar emission \\
     $F_{\nu,{\rm th}}$      & Flux density of dust thermal emission \\
     $F_{\nu,\star}(\lambda)$& Flux density of the star at wavelength $\lambda$ \\
     $G(t-t')$               & Flux density per cross-sectional area from an element at a time $t-t'$ after transit \\
     $h$                     & Specific orbital angular momentum \\
     $H(t-t')$               & Fraction of an element of the ring that is in front of the star at a time $t-t'$ after transit \\
     $i$                     & Inclination of orbit to line of sight \\
     $J(\theta)$             & Ratio of the length of an element along an orbit to its projected size seen by the star \\ 
     $K_j$                   & Fraction of face $j$ of an element that is visible to our line of sight \\
     $L_\star$               & Luminosity of star \\
     $M_\star$               & Mass of star \\
     $N_{\rm f}$             & Number of external faces of an element from which radiation is emitted when optically thick (1 or 4) \\
     $n(\sigma)d\sigma$      & Number of stars in population model with cross-sectional area $\sigma$ to $\sigma+d\sigma$ \\ 
     $p$                     & Probability of witnessing material transit in front of star \\
     $q$                     & Pericentre distance from star \\
     $Q_{\rm abs}(\lambda,D)$& Absorption coefficient at wavelength $\lambda$ for particles of size $D$ \\
     $\langle Q_{\rm abs} \rangle_{D,\star}$ & Absorption coefficient averaged over the particle size distribution and stellar spectrum \\
     $\langle Q_{\rm abs} \rangle_{T}$ & Absorption coefficient averaged over a black body spectrum at temperature $T$ \\
     $Q_{\rm ext}(\lambda,D)$& Extinction coefficient at wavelegnth $\lambda$ for particles of size $D$ \\
     $\langle Q_{\rm ext} \rangle_{D}$ & Extinction coefficient averaged over the particle size distribution \\
     $r$                     & Distance from star \\
     $r_{\rm t}$             & Distance of material from star at point of transit \\
     $R_\star$               & Radius of star \\
     $s$                     & Projected distance along orbit from the point of transit \\
     $ds$                    & Projected length along the orbit of an element of the distribution \\ 
     $S$                     & Total emitting area of an element \\
     $t_{\rm cross}(b=0)$    & Time for material to move $2R_\star$ as seen in projection \\
     $t_{\rm dip}$           & Dip duration \\
     $t_{\rm per}$           & Orbital period \\
     $T_\star$               & Temperature of star \\
     $T(D,r)$                & Temperature of dust of size $D$ at a distance $r$ from the star \\
     $T_{\rm bb}(r)$         & Temperature of black body dust at a distance $r$ from the star \\
     $v_{\rm t}$             & Speed at which material crosses star (tangential velocity) \\
     $\beta$                 & Radiation pressure coefficient; the ratio of the radiation force to that of the star's gravity \\
     $\delta_\lambda$        & Dip depth or dimming at wavelength $\lambda$ \\
     $\Delta$                & Width and height of the distribution of material about the orbit; scales with distance as $\Delta=\kappa r$ \\
     $\Delta_{\rm t}$        & Width and height of the distribution of material about the orbit at the point of transit \\
     $\eta_j$                & Fraction of face $j$ of an element that is blocked by other material along the line of sight \\
     $\theta$                & True anomaly; angular distance around the orbit from pericentre \\
     $\kappa$                & Dimensionless parameter defining width and height of distribution about the orbit (see $\Delta$) \\
     $\lambda$               & Wavelength \\
     $\varpi$                & Orientation of pericentre from line of sight to star \\
     $\dot{\sigma}$          & Rate at which cross-sectional area passes in front of star \\
     $\sigma_{\rm tot}$      & Total cross-sectional area of material \\
     $\sigma(D)$             & Cross-sectional area distribution as a function of particle size $D$ \\
     $\bar{\sigma}(D)dD$     & Fraction of cross-sectioanl area in particles of size $D$ to $D+dD$ \\
     $\tau_{\rm ext}$        & Optical depth due to extinction \\
     $\tau_{\rm abs}$        & Optical depth due to absorption \\
     $\langle \tau_{\rm abs} \rangle_\star$    & Optical depth due to absorption averaged over stellar spectrum \\
     $\phi_j$                & Angle that the normal of face $j$ of an element makes to the line of sight \\
     $\chi$                  & Factor determining dust temperature for optically thick distributions \\ 
     $\Omega_\star$          & Fraction of the observed stellar disk covered by material; covering fraction \\
     \hline
  \end{tabular}
\end{table*}

\bibliographystyle{mnras}
\bibliography{refs.bib} 

\bsp    
\label{lastpage}
\end{document}